\documentclass[preprint,12pt]{elsarticle}
\usepackage{xcolor}

\usepackage{amssymb}
\usepackage{amsmath}
\newcommand{\add}[1]{\textcolor{black}{#1}}

\newcommand{\com}[1]{\textcolor{black}{#1}}
\newcommand{\co}[1]{\textcolor{black}{{#1}}}

 \usepackage{lineno}

\begin{document}



\title{Sound generated by the interaction between shock
and instability waves in supersonic round jets}


\author[1]{Binhong Li} 
\author[1,2]{Benshuai Lyu\corref{mycorrespondingauthor}}
\cortext[mycorrespondingauthor]{Corresponding author}
\ead{b.lyu@pku.edu.cn}

\affiliation[1]{organization={State Key Laboratory of Turbulence and Complex
    Systems, School of Mechanics and Engineering Science, Peking University},
                addressline={5 Yiheyuan Road,
    Haidian District}, 
                city={Beijing},
                postcode={100871}, 
                country={PR China}}

\affiliation[2]{organization={Laoshan Laboratory},
                postcode={266100}, 
                city={Qingdao},
                country={PR China}}

\begin{abstract}
\add{In this paper, we develop an analytical model to investigate the sound
generated by the shock–instability interactions (SII) in supersonic round jets,
extending our previous two‑dimensional planar study to circular configurations.
The jet is represented by a vortex sheet, with its motion modeled by the Euler
equations. Shock and instability waves are modeled using Pack’s approach and the
linear stability theory, respectively, while their interaction is calculated by
solving an inhomogeneous wave equation.} 
\add{Using the Fourier transform and steepest descent method,
we obtain a closed‑form solution for the resulting acoustic field. Results due
to the interaction between the instability waves and one interaction cell
capture the key directivity features of screech reported in experiments and
numerical simulations, indicating that the classic monopole assumption may be
inadequate.} In particular, the screech-tone intensity \add{due to multiple
shock cells} decays rapidly as the observer angle approaches 180 degrees, which
is in better agreement with the experimental data measured by Norum. \add{We
further analyze how the instability wave growth rate influences these
directivity patterns and examine the sound generation efficiency of the
broadband shock-associated noise.} Finally, an examination of near‐field
pressure fluctuations due to the SII reveals that noise is produced primarily
via the Mach wave radiation mechanism.
\end{abstract}

\maketitle



\section{Introduction}
\label{Sec:introduction}
Screech refers to high-intensity acoustic tones in supersonic, shock containing jets~\citep{Raman_review,2019EDINGTON}. Understanding screech is especially important for the design of advanced aircraft since it can cause sonic fatigue for aircraft structures~\citep{Raman_review}. Jet screech was first systematically
studied by Powell in the 1950s. In his pioneering
work,~\citet{1953Powell} proposed the well-established feedback loop theory for screech. In this loop, instability waves originate near the nozzle lip and develop along the jet. When sufficiently amplified, the instability waves interact with the quasi-periodic shock structures and sound is generated. The sound wave propagates in the upstream direction and triggers new instability waves due to the scattering by the nozzle lip, thus completing the feedback loop. The sound source due to the interaction between instability waves and one shock structure was modeled as a monopole by Powell. Assuming constructive interference between several equally-spaced monopole sources in the upstream direction, Powell proposed a formula to predict the screech frequency, i.e.
\begin{equation}
f=\frac{U_{c}}{s(1+M_{c})},
\label{equ_screech f}
\end{equation}
where $s$, $U_{c}$, and $M_{c}$ are the shock spacing, the convection velocity and the convective Mach number of the instability waves, respectively. $M_c$ can be calculated via $M_c=U_c/a_\infty$, where $a_\infty$ is the speed of sound in the free stream. Good agreement between the prediction from Eq.(\ref{equ_screech f}) and the experimental data was observed in rectangular nozzles \citep{Panda_standingwave_111}.

However, in round jets, 
\citet{1953Powell} found that as
the jet pressure increased, the screech frequency experienced abrupt changes. 
This discontinuous change in frequency is commonly referred to as mode staging. Four different modes, i.e. modes A, B, C, and D, were observed by Powell, 
among which mode A can be further divided into two modes named A1 and A2~\citep{M.Merle}. It was found that when the mode transition occurred, the azimuthal structure of both the sound field and instability waves changed simultaneously. 
The axisymmetric mode~{A}{1} and {A}{2} was related to the azimuthal mode $n=0$, and the helical mode C was related to modes $n=\pm1$, while the flapping modes B and D may be decomposed into two equal-strength helical modes with a fixed phase~\citep{2019EDINGTON}. The mode transition from the axisymmetric mode to the helical mode may be explained by LSA. It was found that at low Mach numbers, the total amplification of the axisymmetric mode instability waves is higher than that of the helical mode instability wave, while at high Mach numbers, the total amplification of the helical mode instability wave becomes higher~\citep{LSA_morris}. Considering that the energy of the screech tone is extracted from instability waves, the transition of the most amplified instability mode may explain the transition of the screech mode from A to B~\citep{LST_tam}. However, other mode stagings, i.e. mode A1 to A2 or mode B to C are difficult to be explained by LSA. 

To fully understand the mode staging phenomenon, the role of the guided-jet mode played in the feedback loop has been thoroughly investigated in the last two decades, from the pioneering work by \citet{Tam_and_shen} to that by \citet{unifying_theory}. It was suggested that it is the guided-jet mode, rather than free acoustic wave, that completes the screech feedback loop \citep{nonlinear_in_staging, edgington_modestaging,A1A2modes}. Screech is assumed to result from the interaction between instability waves and shock structures of leading and secondary wavenumbers \citep{unifying_theory,Lixiangru}. The mode staging predicted by the shift of the different guided jet modes that complete the feedback loop showed good agreement with experimental data.

Besides mode staging, jet screech is characterized by its distinct far-field directivity pattern. It was found by \citet{19533Powell} that in rectangular jets, the acoustic wave at the fundamental screech frequency mainly radiates to the upstream direction, while at its first harmonic, the sound emits mainly to the side of the jet ($\psi=90^\circ$). These distinct directivities were subsequently reported by other experiments in both round jets \citep{1983Norm} and rectangular jets \citep{1997_POF_SHWalker, Sources, 2020_sound_sources}, where it was found that at the fundamental frequency, the directivity pattern generally has two lobes, one major lobe to the upstream direction, one secondary lobe to the downstream. At its first harmonic, one major lobe could be found in the directivity pattern near the direction $\psi=90^\circ$, where $\psi$ is the observer angle with respect to the downstream direction. Note that at its first harmonic, the maximal sound radiation angle appears to be not strictly at $90^\circ$, but slightly tilts towards the downstream direction~\citep{numerical_directivity_of_rectangular}. Besides the major lobe near $\psi=90^\circ$, sound emission to the upstream direction could also be significant, as reported by~\citet{1983Norm} and \citet{2020_sound_sources}.

The distinct directivity pattern of screech can be largely explained by the
interference pattern between several monopole sources, as shown by
\citet{1952Powell}. However, it was found that for the fundamental tone,
although the strongest acoustic emission appeared in the upstream direction, a
rapid decay occurred when the observer angle approached
$180^\circ$~\citep{1983Norm}, which could not be predicted by Powell's model. As
a matter of fact, if all the sound sources were monopoles as assumed by Powell
and the frequencies of the fundamental tone were obtained
by~Eq.(\ref{equ_screech f}), the acoustic radiation would be the strongest at
$180^\circ$. The disagreement between the directivity near $180^\circ$ may be
due to the inaccurate assumption of monopoles. To reveal the underlying
mechanism of the distinct directivity of screech, it is important to examine the
shock-instability interaction (SII) in detail instead.

Note that the sound emission due to the SII may \add{be used to explain both
screech and broadband shock-associated noise~(BBSAN)~\citep{broad_band_2013, du1},
with their relation further investigated by \citet{tam_machwave} and
\citet{1986Tam_Proposed}.} One of the earliest studies focused on BBSAN is
\citet{1973Harper}. Based on the phased-array theory, they successfully
predicted the sound frequency generated from \add{the} SII. Subsequently.
\citet{tam_machwave} assumed a semi-analytical form of the disturbances due to
\add{the} SII by multiplying the perturbations induced by shock waves and
instability waves. They found that due to the interaction, some components of
the disturbance may obtain a supersonic phase speed along the jet, which can
therefore lead to Mach wave radiation. The radiation angle and the frequency of
the BBSAN can be readily calculated through the Mach angle relation. Based on
the Mach wave radiation mechanism,~\citet{1986Tam_Proposed} proposed the weakest
link theory suggesting the screech as the limit of the~BBSAN when the observer
angle approaches $180^\circ$. 

Further experimental and theoretical studies on the BBSAN were subsequently conducted by~\citet{1988Tam,1995_annu_tam}. Specifically, Tam developed an analytical model~\citep{1988Tam}, which is formulated by decomposing the flow variables in the governing equations into four components, i.e. the mean flow and perturbations due to instability, shock and acoustic waves resulting from \add{the} SII, respectively. The acoustic wave can be calculated by solving an inhomogeneous wave equation. However, to avoid the extensive numerical calculation when solving the inhomogeneous boundary value problem, Tam proposed an empirical model instead to estimate BBSAN. The parameters needed in this empirical model can be obtained by LSA and experimental measurements~\citep{1988Tam}. The predicted far-field directivity patterns by the empirical model showed satisfactory agreement with the experimental data~\citep{1995_annu_tam}. Recently, \citet{2014TAM} proposed a nonlinear model to explain the generation of harmonics of screech. Instead of examining the whole directivity pattern, they predicted the maximal radiation direction of the screech tones by employing the Mach wave radiation mechanism. The predictions agree well with Norum's experimental data at harmonic frequencies, but appear less so for the fundamental tone.


Besides the Mach wave radiation, the shock leakage is another widely-studied noise generation mechanism for \add{screech}. The oscillation of shock waves induced by instability waves has been extensively reported in early experiments~\citep{1994Suda,S.Kaji}, \add{while the direct experimental observations of the shock leakage mechanism was first proposed by \citet{TAmaning}}. It was then further examined by \citet{2003Lele} and \citet{KS_TAManning_shockleakage}. In the shock leakage mechanism, the shock tip ``leaks" through the jet shear layer and interacts with the instability waves. This interaction leads to the oscillation of the shock tips, which was believed to be the source of the screech. Recent direct experiment observation showed that the upstream-propagating acoustic wave that complete the feedback loop may be generated due to this shock tip oscillation, i.e. the upstream motion of the shock tip~\citep{edgington_shockleakage,Lixiangru}.

As mentioned above, the screech directivity pattern predicted by the monopole-array theory~\citep{1953Powell} and the nonlinear model~\citep{2014TAM} showed satisfactory agreement with the experimental data in round jets \citep{1983Norm}. However, some important
features of the directivity pattern appear not captured satisfactorily by these models, such as the maximal upstream radiation angle. Therefore, an analytical model formulated to describe more realistic physics of the acoustic emission due to \add{the} SII in supersonic round jets is needed. In an earlier work of the authors \citep{li_and_lyu}, an analytical model is developed to calculate the acoustic emission due to \add{the} SII in supersonic two-dimensional jets. However, whether the conclusions drawn in two-dimensional jets remain similar to the round cases is unclear. Following \citet{li_and_lyu}, we aim to develop such an analytical model for supersonic round jets in this paper. 
\com{Note that this model specifically focuses on the sound generation by the
SII. Therefore, in addition to predicting screech, it may also be used to
examine the BBSAN. In particular, this model enables an examination of the sound
generation efficiency associated with the SII and even the BBSAN spectra once the amplitude of the instability waves is known.}

\add{This paper is organized 
as follows. Section
\ref{sec:model} details the analytical derivation of the model. Section
\ref{subsection_directivity} \com{examines} the far-field screech directivity
for both one-cell and multiple-cell interactions. Section \ref{subsec:spectra}
studies the resulting sound generation spectra from \add{the} SII, following which section \ref{sec:near field} examines near-field pressure fluctuations and the noise
generation mechanism. In section \ref{sec:discussion}, 
two-dimensional and
axisymmetric jets are compared, and the role of guided-jet modes is subsequently
addressed. Finally, we summarize our key findings in section \ref{sec:conclusion}.
}

\section{Analytical formulation}
\label{sec:model}
\begin{figure}
   \centering
   \includegraphics[width = 0.5\textwidth]{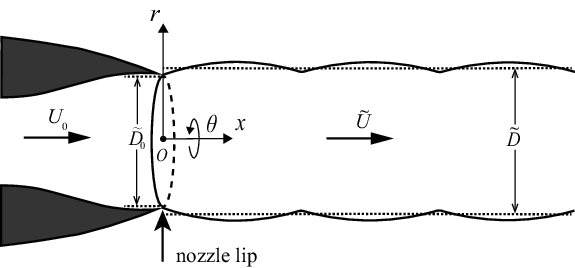}
   \caption{The schematic of the vortex-sheet flow configuration and cylindrical
   coordinates. The origin is fixed at the center of the nozzle while $x $ and
   $r $ represent the streamwise and radial coordinates, respectively.}
\label{fig:example1}
\end{figure}
We assume that the jet is of the vortex-sheet type. As shown
in Fig.~\ref{fig:example1}, the origin of the cylindrical coordinate $(r, x, \theta)$ is located at the 
center of the nozzle exit. $\tilde{D}$ is the diameter of the fully expanded jet. Note that $\tilde{D}$ is generally not equal to the diameter of the nozzle, i.e. $\tilde{D}_0$. Instead, we take the diameter of the fully expanded jet as the base flow diameter~\citep{tam_diameter}. 
 $\tilde{U}$ is
the velocity of the fully expanded jet after exiting from the nozzle. To simplify the problem, we non-dimensionalize velocities and lengths by $\tilde{U}$ and $\tilde{D}$, respectively. In addition, densities and temperatures are non-dimensionalized by the mean density $\tilde{\rho}_{0+}$ and the total temperature $\tilde{T}_{0+}$ outside the jet, respectively. Note that the subscript ``$+$'' represents the parameters outside the jet. We use the symbols with a tilde to represent dimensional variables, while those without denote non-dimensional variables. For example, $U_{0}$ is the non-dimensional jet velocity at the nozzle exit plane, and the fully expanded base flow illustrated in Fig.~\ref{fig:example1} takes the form: $ \boldsymbol{u}_{0}= 0$ when $r >1/2$, while $ \boldsymbol{u}_{0}=  \mathbf{{e}}_{x }$ when $r \leq1/2$.
Here $\mathbf{e}_{x }$ represents the unit vector in the $x $ direction.

Considering that the Reynolds number is high in practical applications and the viscosity is expected to have negligible effects, we start from the Euler equations shown as follows
   %
\begin{subequations}
\begin{align}
    \frac{\mathrm{D} \rho}{\mathrm{D} t  } + \rho\boldsymbol{\mathbf{\nabla} \cdot }\boldsymbol{u}=0 &,\label{equ:continunity}\\
    \rho \frac{\mathrm{D}\boldsymbol{u}}{\mathrm{D} t }= -\boldsymbol{\nabla} p &,\label{equ:momentum}\\\label{equ:entropy} 
     \frac{\mathrm{D} s}{\mathrm{D} t } = 0&, 
 \end{align}
\end{subequations}
where $t$ represents time, $\boldsymbol{u}=(u,v,w)$ the velocity, $p$ the
pressure, $\rho$ the density, $s$ the entropy, and $ {\mathrm{D}}/{\mathrm{D} t  }$ denotes ${\partial }/{\partial t}+\boldsymbol{u\cdot\mathbf{\nabla}}$. Note that we use the
isentropic condition here, i.e. Eq.(\ref{equ:entropy}), because the entropy increase
across weak shock is a high-order small term~\citep{1995Kerchen}.

Following \citet{1988Tam}, we decompose the variables in Eqs.(\ref{equ:continunity})-(\ref{equ:entropy}) into the mean and perturbation components induced by the shock, instability, and acoustic waves due to the SII, respectively. For example, with the decomposition, the velocity can be written as  
\begin{equation}
   \boldsymbol{u}=\boldsymbol{u}_{0}
   +\delta \boldsymbol{u}_{m}
   +\epsilon\boldsymbol{u}_{v}
   +\delta^{2}\boldsymbol{u}_{m2}
   +\epsilon^{2}\boldsymbol{u}_{v2}
   +\delta\epsilon\boldsymbol{u}_{i}+...\ ,
   \label{equ:decomp_u}
\end{equation}
where the subscripts $0$, $m$, $v$, and $i$ in Eq.(\ref{equ:decomp_u}) represent the mean flow and the perturbations due to shock, instability, and SII-generated acoustic waves, respectively. Such definitions of the subscripts remain the same in the rest of this paper. $\delta$ and $\epsilon$ in Eq.(\ref{equ:decomp_u}) denote
the strength of the shock and instability waves, respectively. These two
parameters are assumed to be of a small magnitude representing small
perturbation compared to the mean flow, that is, $\delta\ll 1$ and $\epsilon \ll 1$.
 The second-order terms $\delta^2$ and $\epsilon^2$ can therefore be neglected. However, 
the $\delta \epsilon$ term is retained and represents the SII-generated acoustic waves \citep{1995Kerchen,li_and_lyu}.

The temperature across the jet  can be
different, while the mean pressure $p_{0}$ remains constant and identical in both regions \citep{LSA_morris}. If a perfect gas is assumed, then ${c}_{0\pm}=\sqrt{\gamma { p}_{0\pm}/{\rho_{0\pm}}}$  can be used to calculate the speed of
sound inside and outside the jet, respectively. Here $\gamma$ denotes the specific heat ratio. Note that the subscript ``$-$'' represents the parameters inside the jet. We define $M_{-}=1/{c}_{0-}$ and $M_{+}=1/{c}_{0+}$ to denote the jet Mach number based on the speed of sound inside and outside the jet, respectively.  Substituting the expansions to the Euler equations and collecting the terms $O(\delta)$, $O(\epsilon)$, and
$O(\delta\epsilon)$, we obtain the equations governing the shock, instability, and
their interaction, respectively.

Regarding the instability waves, collecting the $O(\epsilon)$ term results in the well-known convective-wave equations \citep{196batchelor}:
\begin{equation}  
    \mathbf{\nabla}^{2}\phi_{v\pm}-M_{\pm}^2\frac{\mathrm{D}^{2}\phi_{v\pm}}{\mathrm{D} t^{2}}=0,
\end{equation}
where $\phi_{v+}$ and $\phi_{v-}$ denote the velocity potentials for the instability waves outside and inside the jet, respectively.
Using the dynamic and kinematic boundary conditions imposed on $r=1/2$, we obtain the dispersion relation between the streamwise wavenumber $\alpha$ and the angular frequency $\omega$:
\begin{equation}
    \frac{(\omega-\alpha)^{2}\mu_{+}}{M_{+}^{2}}\frac{H_{n}^\prime (\mu_{+}/2)}{H_{n}(\mu_{+} /2)}-\frac{\omega^{2}\mu_{-}}{M_{-}^{2}}\frac{J_{n}^\prime (\mu_{-}/2)}{J_{n}(\mu_{-}/2)}=0,
    \label{equ:despersion relation}
\end{equation}
where $n$ represents the azimuthal mode number, $\mu_+$ and $\mu_-$ are respectively defined by $\mu_{+}=\sqrt{\omega^{2}M_{+}^{2}-\alpha^{2}}$ and $
\mu_{-}=\sqrt{M_{-}^{2}(\omega-\alpha)^{2}-\alpha^{2}}$, the branch cuts of which are properly chosen to make sure that the imaginary part of $\mu_{+}$ is positive, $J_{n}$ and $H_{n}$ denote the $n$th-order Bessel and Hankel functions of the first kind, respectively, and the superscript $^\prime$ means taking the first derivative of the underlying function. 
The instability-induced velocity, pressure, and vortex sheet's displacement perturbations can be obtained from the potential function, all of which are shown in Appendix \ref{appendix1}.

By collecting the $O(\delta)$ terms, we find that the weak shock can be modeled by the linearized wave equation:
\begin{equation}
    \mathbf{\nabla}^{2}\phi_{m}-M_{-}^{2}\frac{\partial^{2}\phi_{m}}{\partial x^{2}}=0.
    \label{equ:waveequation}
\end{equation}
Pack's model is widely used to describe the velocity potential~\citep{1950Pack}, i.e.
\begin{equation}
    \phi_m=\sum_{j=1}\mathcal{A}_{j}J_{0}(2\zeta_{j}r)\sin(2\zeta_{j}x/\beta),
    \label{Pack's model}    
\end{equation}
where $\mathcal{A}_{j}=-2\beta{(1-U_{0})}/{\zeta_{j}^{2}J_{1}(\zeta_{j})}$,
$\beta=\sqrt{M_{-}^{2}-1}$, $U_0$ the streamwise velocity at the nozzle exit, $J_{0}$ and $J_{1}$ are $0$th- and $1$st-order Bessel functions
of the first kind respectively, and $\zeta_j$ denotes the $j$th zero of the $J_0$ function.
In the present study, we primarily focus on the
leading-order mode. Higher-order modes can be easily included at a later stage
should necessity arise. In what follows the subscripts in parameters $\mathcal{A}_{1}$ and $\zeta_{1}$ are  omitted  for  clarity. 

For the leading-order mode, the shock wave distribution is a 
periodic function of $x$ with a shock spacing 
$s=\pi\beta/\zeta.$
The corresponding velocity, pressure, and vortex sheet's displacement perturbations due to shock structures can be obtained from the potential function and are shown in Appendix A. 

Having obtained the shock and instability waves, we are now in a position to
consider the perturbation induced by the SII. Following \citet{li_and_lyu}, the corresponding velocity potential $\phi_{i}$ satisfies the following inhomogeneous wave equation
  \begin{equation}
      \mathbf{\nabla}^{2}\phi_{i}-M^2\frac{\mathrm{D}^{2}\phi_{i}}{\mathrm{D} t^2}=\frac{-1}{M^2}[2(\boldsymbol{u}_m\boldsymbol{\cdot}\boldsymbol{\nabla} p_{v}+\boldsymbol{u}_v\boldsymbol{\cdot}\boldsymbol{\nabla} p_{m})+(\gamma-1)(p_{m}\boldsymbol{\nabla}\boldsymbol{\cdot}\boldsymbol{u}_{v}+p_{v}\boldsymbol{\nabla}\boldsymbol{\cdot}\boldsymbol{u}_{m})].
      \label{equ_sound_inhomogenous}
        \end{equation}
 The source term in Eq.(\ref{equ_sound_inhomogenous}) differs inside and outside the jet, therefore, the corresponding $\phi_{i\pm}$ satisfy different governing equations (see Eqs.(\ref{equ_homogenous}) and (\ref{equ_sound_inhomogenous2})). Additionally, the total pressure and total displacement should be continuous across the vortex sheet (see Eqs.(\ref{equ:boundary condition 1}) and (\ref{equ:boundary condition 2})). Full details of these governing equations and boundary conditions can be found in Appendix B.
 
To obtain the solution to Eq.(\ref{equ_homogenous}) and Eq.(\ref{equ_sound_inhomogenous2}) subject to boundary conditions Eq.(\ref{equ:boundary condition 1}) and Eq.(\ref{equ:boundary condition 2}), the Fourier transform is used. The Fourier transforms $G_{\pm}(\lambda,r)$ are defined as $G_{\pm}(\lambda,r)=\int_{-\infty}^{+\infty} \phi_{i\pm}(x,r)\mathrm{e}^{\mathrm{i} \lambda x}\mathrm{d} x$,
where $\lambda$ denotes the wavenumber in the streamwise direction.
Outside the jet, it can be found that $G$
satisfies $ G_{+}(\lambda,r)= D_{+}(\lambda)H_{n}(\gamma_{+} r),$
 where $\gamma_{+}(\lambda)=\sqrt{\omega^{2}M_{+}^{2}-\lambda^{2}}$
and $D_{+}$ is an undetermined function of
$\lambda$.
    	\begin{figure}
		\centering
		\includegraphics[width = 0.55\textwidth]{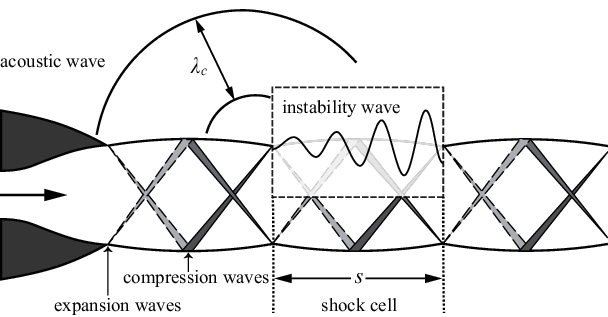}
		\caption{Schematic of a source term located within one single
		shock cell. The Prandtl-Meyer
		expansion waves occur at the nozzle lip and are reflected as compression waves (oblique shock waves) on the opposite jet boundary, thereby forming one shock structure \citep{shock_structure1,Mehta23}. The acoustic wave (with a wavelength $\lambda_c$) is generated through the interaction between shock and instablity waves. The total field is
		equivalent to a linear superimposition of the results from all
		shock cells.}\label{fig:schematic}
	\end{figure}

When Fourier transform is applied to the source term of~Eq.(\ref{equ_sound_inhomogenous2}), following \citet{li_and_lyu},
we can
simplify the problem by noting the periodicity of the function $\cos(2\zeta x/\beta)$
and  $\sin(2\zeta x/\beta)$. This means that we
only need to consider the interaction within one shock cell, as shown in
Fig.~\ref{fig:schematic}, and the total interaction field would be a simple
linear combination of several shock cells. Mathematically, this 
procedure can be justified by the linearity of boundary conditions~Eq.(\ref{equ:boundary condition 1}) and~Eq.(\ref{equ:boundary condition 2}), and the governing equation~(\ref{equ_sound_inhomogenous2}). Physically, this corresponds to the effective sound generated by the interaction between the
instability waves and one single shock cell. However, we emphasize that by doing so we do not imply that such an effective source is physically localized. The source term on the right side of Eq.(\ref{equ_sound_inhomogenous}) has a periodic nature by construction, but the bounds of integration may be across one or several shock cells due to the linearity of Eq.(\ref{equ_sound_inhomogenous}). The main purpose of this decomposition is to examine the sound generation of the one-cell interaction, which was assumed to be of the monopole type by Powell.
Upon limiting the integration interval to be within one shock cell, we can define $ \mathcal{I}_s(\lambda)={2a}/{[(\alpha+\lambda)^2-a^2]}\sinh\left((\alpha+\lambda){\pi}/{a}\right)\mathrm{e}^{-\mathrm{i}(\alpha+\lambda)\frac{\pi}{a}}$ and $\mathcal{I}_c(\lambda)={-2\mathrm{i}(\alpha+\lambda)}/{[(\alpha+\lambda)^2-a^2]}\sinh\left((\alpha+\lambda){\pi}/{a}\right)\mathrm{e}^{-\mathrm{i}(\alpha+\lambda)\frac{\pi}{a}}$, with $a$ defined by $a=2\zeta/\beta$.

The function
$G_{-}(\lambda,r)$ can be divided into two parts, a particular solution,
$G^{p}(\lambda,r)$,
and a complementary solution, $G^{c}(\lambda,r)$.   
    The particular solution is obtained in Appendix \ref{appendixb},
while the complementary solution can be written as $G^{c}(\lambda,r)=D_{-}(\lambda)J_{n}(\gamma_{-}r)$,
    where
    $\gamma_{-}=\beta\sqrt{(\lambda-M_{1})(\lambda-M_{2})}$, $D_{-}$ an undetermined function of $\lambda$, and the two coefficients
$M_1$ and $M_2$ are  respectively defined by $M_{1}={-M_{-}\omega}/(M_{-}+1)$ and  $M_{2}={-M_{-}\omega}/(M_{-}-1).$

   By	applying
 the  Fourier transform to the two boundary conditions Eq.(\ref{equ:boundary condition 1}) and Eq.(\ref{equ:boundary condition 2}), and using the solution forms of $G_\pm(\lambda,r)$,  we can solve $D_+(\lambda)$ and $D_-(\lambda)$ and obtain 
\begin{subequations}
\begin{equation}
    \begin{aligned}
    D_{+}(\lambda)=\frac{1}{\eta}\bigg[\mathcal{X}_0\mathcal{I}_{c}\gamma_{-}J^\prime _{n}\left(\frac{\gamma_{-}}{2}\right)-\frac{M_{-}^{2}}{M_{+}^{2}}(\omega+\lambda)\bigg((\mathcal{S}_1\mathcal{I}_{s}+&\mathcal{S}_2\mathcal{I}_{c}+G^{p^\prime})J_{n}\left(\frac{\gamma_{-}}{2}\right)\\
    &-\gamma_{-}G^{p}J^\prime _{n}\left(\frac{\gamma_{-}}{2}\right)\bigg)\bigg],
    \end{aligned}
\end{equation}
\begin{equation}
    \begin{aligned}
    D_{-}(\lambda)=\frac{1}{\eta}\bigg[\left(1+\frac{\lambda}{\omega}\right)\bigg(\mathcal{X}_0\mathcal{I}_c+(\omega+\lambda)&\frac{M_{-}^2}{M_{+}^2}G^{p}\bigg)\gamma_{+}H_n^\prime \left(\frac{\gamma_{+}}{2}\right)\\
    -&\omega(\mathcal{S}_1\mathcal{I}_{s}+\mathcal{S}_2\mathcal{I}_{c}+G^{p^\prime}) H_n\left(\frac{\gamma_{+}}{2}\right)\bigg],
    \end{aligned}
\end{equation}    
\end{subequations}
where 
\begin{equation}
    \eta(\lambda)=\omega\gamma_{-}H_{n}\left(\frac{\gamma_{+}}{2}\right)J^\prime _{n}\left(\frac{\gamma_{-}}{2}\right)-\frac{M_{-}^{2}}{M_{+}^{2}}\frac{(\omega+\lambda)^{2}}{\omega}\gamma_{+}H^\prime _{n}\left(\frac{\gamma_{+}}{2}\right)J_{n}\left(\frac{\gamma_{-}}{2}\right).
\end{equation}	
The inverse Fourier transform of $D_{+}(\lambda)$ yields $\phi_{i+}$, i.e.
\begin{equation}
        \phi_{i+}(x,r)=\frac{1}{2\pi}\int_{-\infty}^{+\infty}D_{+}(\lambda)H_{n}(\gamma_{+}r)\mathrm{e}^{-\mathrm{i}\lambda x}\mathrm{d}\lambda.
        \label{equ:numerical integration}
\end{equation}
\begin{figure}
		\centering	
		\includegraphics[width = 0.6\textwidth]{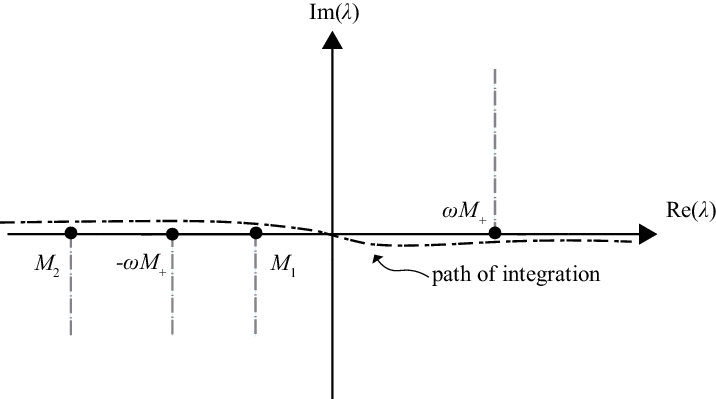}
		\caption{The branch points, branch cuts, and integral path in the complex $\lambda$ plane. 
		}\label{fig:example3}
	\end{figure}
It is straightforward to verify that $\lambda=-\alpha$ is a simple zero for
$\eta(\lambda)$ according to the dispersion relation Eq.(\ref{equ:despersion
relation}).
Besides, $\lambda=-\alpha$ is a simple                     
zero for $\mathcal{I}_s(\lambda)$
and a second-order zero for $\mathcal{I}_c(\lambda)$. In addition, $G^p$ is proportional to $\mathcal{I}_s(\lambda)$ and $\mathcal{I}_c(\lambda)$ according to Eq.(\ref{equ:G_p}). Therefore, $\lambda=-\alpha$ is not a pole for $D_{+}(\lambda)$. Therefore, we do not need to deform the integration path to wrap around  $\lambda=-\alpha$ in the complex $\lambda$ plane while performing the inverse Fourier transform Eq.(\ref{equ:numerical integration}) \citep{Briggs}.

As illustrated in Fig.~\ref{fig:example3}, the integration path in Eq.(\ref{equ:numerical integration}) goes from $-\infty+\mathrm{i}\sigma$ to $+\infty-\mathrm{i}\sigma$ in the complex $\lambda$ plane and crosses the real $\lambda$ axis at $\lambda=0$. Here $\sigma$ is a small real parameter. By doing so, we can ensure that $|\gamma_{+}|>\sigma$ along the integration path. Because the imaginary part of $\gamma_{+}$ should be positive when $|\lambda|\rightarrow \infty$ (so that $H_n(\gamma_+)\rightarrow 0$ when $|\lambda|\rightarrow \infty$), the branch
cuts of  $\gamma_{+}$  at branch points $\lambda=\pm \omega M_{+}$ are chosen to extend to the upper and lower half planes, respectively. With this choice of branch cuts, $0<\mathrm{arg}(\gamma_{+})<\pi/2$ along the integration path. The branch points of $\gamma_{-}$, i.e. $M_{1}$ and $M_{2}$ are on the
negative real $\lambda$ axis; the branch cuts of $\gamma_{-}$ are chosen to extend down to the lower half plane so as not to cross the integration path. 

	\begin{figure}
		\centering	
		\includegraphics[width = 0.55\textwidth]{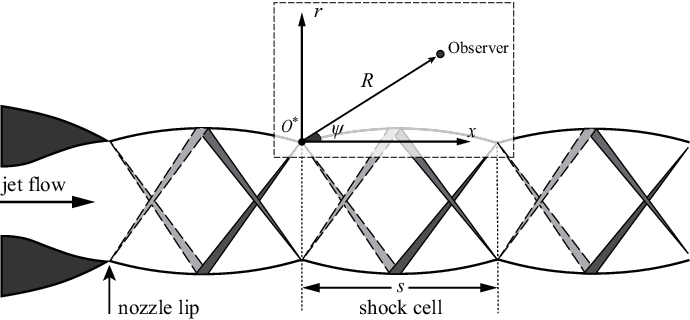}
		\caption{Schematic of one azimuthal plane of the jet. $R$ is the distance from the observer to the starting point of one shock cell and $\psi$ represents the observer angle with respect to the downstream direction. In this plane, $r=R\sin\psi$ and $x=R\cos \psi$. }\label{fig:axis}
	\end{figure}	

Note that in the far field, $H_{n}(\gamma_{+}r)$ takes an approximation form, i.e. $\sqrt{\frac{2}{\pi\gamma_{+}r}}\mathrm{e}^{\mathrm{i}(\gamma_{+}r-n\pi/2-\pi/4)}$. Therefore, when $r\gg 1/\sigma$, $ \phi_{i+}(x,r)$ can be evaluated as
\begin{equation}
\phi_{i+}(x,r) \approx \frac{1}{2\pi} \int_{-\infty}^{+\infty} D_+(\lambda) \sqrt{\frac{2}{\pi \gamma_+ r}} \, \mathrm{e}^{-\mathrm{i} \left( \lambda x - \gamma_+ r + \frac{n\pi}{2} + \frac{\pi}{4} \right)} \, \mathrm{d}\lambda.
    \label{equ:far-field estimation}
\end{equation}
Assume that the observer is in one specific azimuthal plane as illustrated in Fig.~\ref{fig:axis}. We construct a new coordinate centered at the starting point of one shock cell (the effective source region). The distance between the observer and the origin $O^{*}$ is $R$ and the observer angle with respect to the downstream direction is denoted by $\psi$.  Eq.(\ref{equ:far-field estimation}) can be then approximated in terms of $R$ and $\psi$ as
\begin{equation}
\phi_{i+}(x,r) \approx \frac{1}{2\pi} \int_{-\infty}^{+\infty} D_+(\lambda) \sqrt{\frac{2}{\pi \gamma_+ R \sin\psi}} \, \mathrm{e}^{-\mathrm{i} \left( \lambda R \cos\psi - \gamma_+ R \sin\psi - \frac{\gamma_+}{2} + \frac{n\pi}{2} + \frac{\pi}{4} \right)} \, \mathrm{d}\lambda.
\end{equation}
Using the
steepest descent method and noting that the saddle point is $\lambda=-M_{+}\omega \cos\psi$, we can express $\phi_{i+}$ as a function of distance $R$ and observer angle
$\psi$ in the far field, i.e.
    \begin{equation}
       \phi_{i+}(r, \theta) \approx \frac{1}{R} D_+ \left( - M_+ \omega \cos\psi \right) \, \mathrm{e}^{\mathrm{i} \omega \left( M_+ R + M_+ \sin\psi - \frac{\pi}{4} \right) - \mathrm{i} \left( \frac{n}{2} - \frac{1}{4} \right) \pi} + O(R^{-2}).
        \label{steepest_decent_way}
    \end{equation}
so the corresponding pressure takes the form 
    \begin{equation}
       p_{+}(r, \theta) \approx \frac{-\mathrm{i}\omega}{R} D_+ \left( - M_+ \omega \cos\psi \right) \, \mathrm{e}^{\mathrm{i} \omega \left( M_+ R + M_+ \sin\psi - \frac{\pi}{4} \right) - \mathrm{i} \left( \frac{n}{2} - \frac{1}{4} \right) \pi} + O(R^{-2}).
        \label{steepest_decent_way_P}
    \end{equation}
    
    \section{Results} 
\label{section:results and discussion}
\add{In what follows, we start by using Powell's model to predict the screech
frequency. The predicted frequency $\omega$ is then substituted into the dispersion
relation Eq.(\ref{equ:despersion relation}) to determine the corresponding
streamwise wavenumber $\alpha$. 
Finally, both $\omega$ and $\alpha$ are substituded into 
Eq.(\ref{steepest_decent_way_P}) to calculate the associated far-field sound
pressure.  To validate the far-field approximation, 
sound propagating towards the observer angle of
$\psi=90^\circ$ is evaluated via numerical integration. Key results, including
the predicted directivity patterns, sound generation efficiency spectra of the BBSAN, and
near-field pressure fluctuations, are shown in the following sections.} 

We use Powell's formula Eq.(\ref{equ_screech f}) to predict the screech frequency. The shock spacing $s$ can be calculated using Pack's model Eq.(\ref{Pack's model}), i.e.
\begin{equation}
    s=\frac{\pi}{2.4048}\sqrt{M_{-}^2-1}.
    \label{equ:shock_spacing}
\end{equation} 
The convection velocity $U_c$ is assumed to be proportional to the velocity of the fully expanded jet flow, i.e.
\begin{equation}
    U_c=\kappa,
    \label{equ:Uc}
\end{equation}
where $\kappa$ takes the value of 0.7 in this study \citep{1973Harper}. The convective Mach number can be calculated using Crocco-Busemannn's rule,
\begin{equation}
    M_{c}=\frac{\kappa M_{-}}{\sqrt{1+\frac{\gamma-1}{2}M_{-}^2}}.
    \label{equ:convective velocity}
\end{equation}
 Combining Eq.(\ref{equ_screech f}), Eq.(\ref{equ:shock_spacing}), Eq.(\ref{equ:Uc}), and Eq.(\ref{equ:convective velocity}) yields
\begin{equation}
    f=\frac{1.6834}{\pi}(M_-^2-1)^{-1/2}\left[1+0.7\left(1+\dfrac{\gamma-1}{2}M_{-}^2\right)\right]^{-1}.
    \label{equ:screech_fre}
\end{equation}
The Sound Pressure Level (SPL) is defined by
\begin{equation}
    \mathrm{SPL}=10\log_{10}\left(\frac{P}{P_{\mathrm{ref}}}\right)^{2},
    \label{equ:SPL}
\end{equation}
where the reference pressure is set as $P_\mathrm{ref}=2\times 10^{-5} $.

Before being used to study the directivity of the resulting sound, Eq.(\ref{steepest_decent_way_P}) is validated by numerically integrating~Eq.(\ref{equ:numerical integration}) when $\psi=90^\circ$. Excellent agreement is achieved between the prediction using~Eq.(\ref{steepest_decent_way_P}) and~Eq.(\ref{equ:numerical integration}) in the far field (results omitted here for brevity). Specifically, when $R$ is beyond $9$, the difference between the two approaches is within 1 dB, and when $R\geq 20 $, the difference further reduces to 0.1 dB. These findings indicate that $kR\approx6.19$ may be used to approximately separate the near and far fields.

 
\subsection{Directivity of the sound field}
\label{subsection_directivity}
In the pioneering work of Powell, the sound source generated by the interaction between the instability waves and one shock cell was modeled as a monopole. Therefore, the distinct directivity patterns of the screech are purely due to the interference between several equally-spaced monopole sources. To verify this assumption, Sec.\ref{subsub:directivity} compares the directivity patterns predicted by both our model and Powell’s model for a single‐cell interaction and for interactions involving several equally spaced cells. The experimental data reported by Norum and numerical result reported by Tam are included for comparison. Following the comparison, Sec.\ref{subsub:effects} focuses on the spatial growth rate of the instability waves and assesses how variations in this growth rate influence the directivity pattern in our model.

 \subsubsection{Directivity pattern of the the screech}\label{sec:one-interaction}
 \label{subsub:directivity}
 \begin{figure}
	\centering
	\includegraphics[width = 0.45\textwidth]{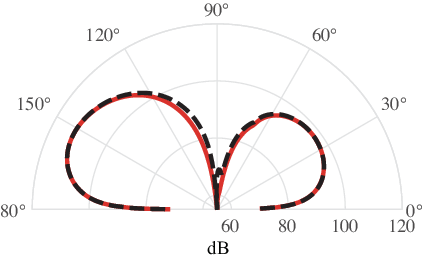}
	\caption{The directivity patterns when three shock-cell interactions are included. The Mach number of the fully expanded jet flow is 1.19. The relative intensity of the three sources are (0.5, 1, 0.5) and (0.2, 1, 0.8) for the red solid line and the black dashed line, respectively.}
		\label{changeIntensity}
	\end{figure}

The directivity patterns due to one-cell interaction can be directly calculated
via Eq.(\ref{steepest_decent_way_P}). Regarding several cell interactions, the
directivity patterns are constructed by incorporating spatial phase differences
via, i.e.
$\mathrm{e}^{\mathrm{i}m\left(\lambda_{0}+\mathrm{Re}(\alpha)\right)2\pi/a}$,
where $\lambda_{0} = -\omega M_{+}\cos\psi$ and $m$ represent an integer taking
the value of $0, \pm1, \pm2...$ depending on how many shock cells are included.

Before comparing the prediction with the experimental and numerical data, we
first discuss the number of sources used and their relative intensities. Based on
Powell's model~\citep{1953Powell},~\citet{1983Norm} used nine equally-spaced
monopole sources with a parabolic intensity distribution. Similar methods were
also used by~\citet{numerical_directivity_of_rectangular}, where six sources
with a relative intensity of (0.5, 0.5, 1, 0.5, 0.25, 0.25) were included. The
predicted directivity patterns due to these monopole sources were respectively
compared with the experimental data~\citep{1983Norm} and numerical simulation
results~\citep{numerical_directivity_of_rectangular}.  The number of sources
included in these studies, however, was uncharacteristically large, because
experimental data showed that the effective source region of screech is likely
to be within $2$ or $3$ shock
structures~\citep{Panda_standingwave,2014TAM,2017_jfm_sources,Sources}. In this
paper, instead of using 6 or 9 monopoles, we choose to use only three effective
sources based on these experimental observations.

Regarding the relative intensity of the three sources, considering that the
energy of the screech tone is extracted from the instability waves, the
intensity of the screech source may be directly related to that of the
instability waves. Both experimental measurements~\citep{1992Powell} and numerical simulations~\citep{hanshuaibin} report that the intensity profile of instability waves along the jet follows an approximately “parabolic” form (i.e., initially increases, reaches a peak, and then decays). Linear stability analyses using the Parabolized Stability Equations (PSE) reproduce such growth‑saturation‑decay behaviour, which qualitatively resembles a parabola~\citep{PSE_herbert, my_pse, my_pse_2}. 
Therefore, we choose a
parabolic distribution such as $(0.5, 1, 0.5)$ in order to predict the
directivity.

In fact, as shown by~\citet{M.Kandula}, the impact of varying source strengths was unimportant with regard to the principal lobe, while only noticeable in the secondary or minor lobes. To put this into perspective, the directivity patterns obtained by changing the relative strengths of the three sources are plotted in Fig.~\ref{changeIntensity}, where relative intensities of (0.5, 1, 0.5) and (0.2, 1, 0.8) are assumed, respectively. From Fig.~\ref{changeIntensity}, it is evident that the two results are virtually the same apart from the marginal difference in terms of lobe width. Thus, in what follows, we choose to use the relative intensity of (0.5, 1, 0.5).

We start by comparing this model prediction and Powell's model prediction with the experimental data reported by \citet{1983Norm}.  The Mach number of the fully expanded jet flow is chosen to be 1.19 and the B mode is examined in accordance with the operation conditions used by \citet{1983Norm}. The azimuthal mode of the instability wave is therefore $n=1$ in this model. 
  \begin{figure}
	\centering
	\includegraphics[width = 0.75\textwidth]{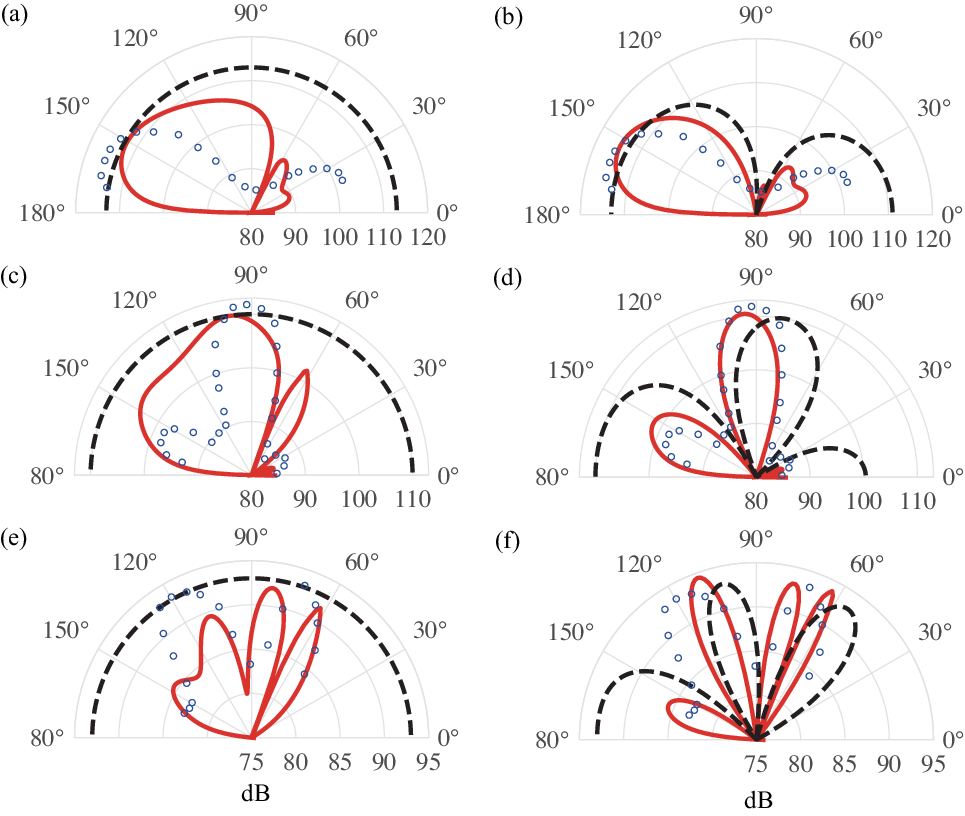}
	\caption{Comparisons of the far-field directivity patterns obtained by our model, by Powell's model, and experimental data~\citep{1983Norm}. The panels $(a, c, e)$ are for one-cell interaction, while $(b, d, f)$ are for three-cell interactions. Panels $(a, b)$ the fundamental frequency; $(c, d)$ the first harmonic; $(e, f)$ the second harmonic. The red solid lines, the black dashed lines, and the blue circles represent this model's prediction, Powell's model prediction, and the experimental data, respectively. The Mach number of the fully expanded jet flow is 1.19.  }
		\label{1.19_1}
	\end{figure}

As shown in Fig. \ref{1.19_1}(a, c, d), 
Powell's model consistently predicts uniform circular patterns due to its monopole assumption. Therefore, when the screech source from one-cell interaction is regarded as a monopole, the unique directivity observed in experiments results from the constructive interference between several equally-spaced monopoles.  In contrast, the present model predicts distinct directional lobes. 

Regarding the fundamental tone, as shown in Fig.\ref{1.19_1}(a), the present model predicts a major lobe in the upstream direction and a weaker lobe
in the downstream direction. This qualitatively matches the experimental two-lobe structure (dominant upstream, weaker downstream), although the lobe's relative intensity and width differ.  This is not surprising because we only examine the directivity pattern due to one-cell interaction.  When the three sources are combined in Fig.\ref{1.19_1}(b),  Powell's model predicts two lobes, which are of nearly equal intensity and peak at $180^\circ$ and $0^\circ$, respectively. We can see that when $\psi$ approaches $180^\circ$, the monopole array theory predicts increasingly large noise radiation, which clearly contradicts the experiment. On the other hand, the present model predicts a similar major lobe in the upstream direction, and a weaker lobe in the downstream direction. The relative intensity and width of the two lobes agree better with the experimental data than the monopole array theory. More importantly, the present model correctly captures the radiation decay near $180^\circ$. However, the relative intensity of the two lobes does not precisely match the experimental results. Considering the many assumptions made in the model, such deviations may be inevitable.

Regarding the first harmonic, as shown in Fig.\ref{1.19_1}(b), the present model
predicts a major side lobe ($90^\circ$), a minor upstream lobe, and another lobe
near $60^\circ$ with a relatively lower intensity. The experimental data is
somewhat different. Two evident lobes can be found. The dominant one points to
the side of the jet ($90^\circ$), while the secondary one points to the upstream
direction. These two lobes are more well-separated than those in the prediction.
In addition, a much smaller lobe appears in the downstream direction, in
contrast to that pointing to $60^\circ$ in the prediction. Despite differences,
overall similarity exists between the prediction and the experimental results.
When three sources are combined, the prediction from the monopole array theory
shows three lobes. Two major lobes peak at $80^\circ$ and $180^\circ$,
respectively, which are of nearly the same intensity. One minor lobe in the
downstream direction peaks at $0^\circ$. The lobe position, width and relative
intensity of the three lobes agree unsatisfactorily with the experimental data.
In particular, the quick decay as the observer angle approaches $180^\circ$ is
not captured. In contrast, we can see that the present model's prediction agrees
well with the experimental data in terms of lobe position, width and relevant
intensity. In particular, the radiation decay as $\psi$ approaches $180^\circ$
is successfully captured by this model. 

With regard to the second harmonic, the present model predicts four lobes: two major lobes between $60^\circ<\psi<90^\circ$, one major lobe near $110^\circ$, and a minor upstream lobe. Experiments show two major lobes (slightly upstream and downstream) and a weak upstream lobe, with major radiation ranges aligning broadly with predictions. After combining three sound sources, Powell's model predicts three equal-intensity lobes $(\sim50^\circ, 100^\circ, 180^\circ)$, whose positions appear to differ from those in experiments. On the other hand, the present model predicts four lobes. Because of limited angle resolution, it is difficult to quantify the exact number of lobes in the experiments. Nevertheless, the lobe position and the overall shape of the directivity patterns are in better agreement with the experimental data than Powell's model.
	
\begin{figure}
	\centering
	\includegraphics[width = 0.75\textwidth]{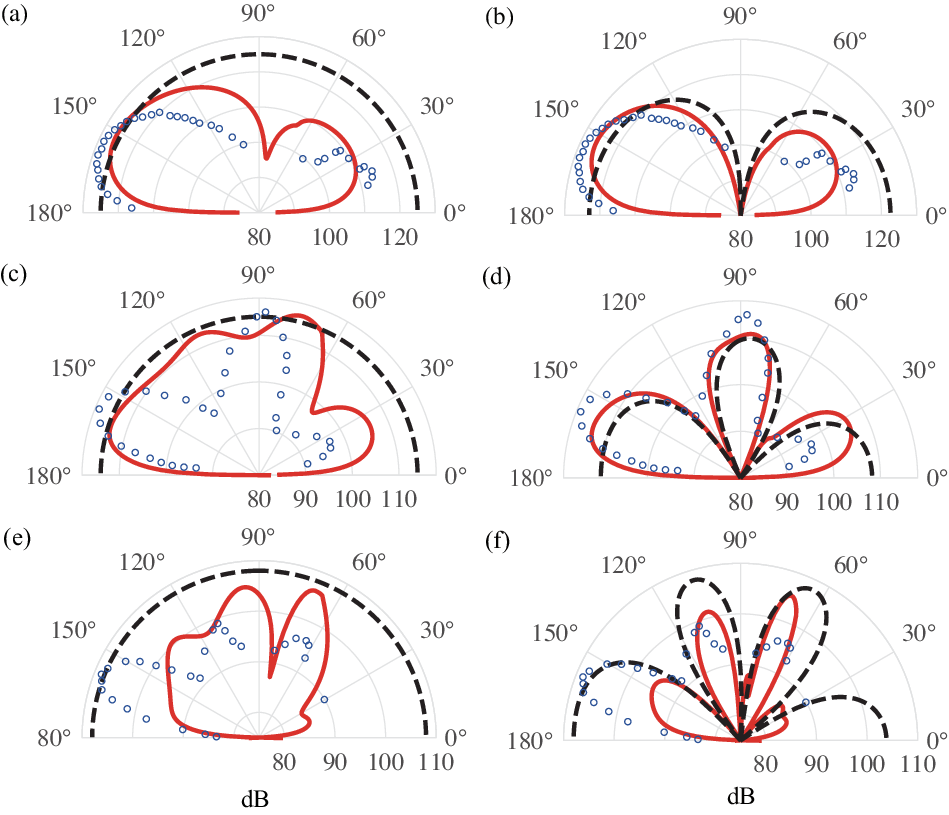}
	\caption{Comparisons of the far-field directivity patterns obtained by our model, by Powell's model, and experimental data~\citep{1983Norm}. The panels $(a, c, e)$ are for one-cell interaction, while $(b, d, f)$ are for three-cell interactions. Panels $(a, b)$ the fundamental frequency; $(c, d)$ the first harmonic; $(e, f)$ the second harmonic. The red solid lines, the black dashed lines, and the blue circles represent this model's prediction, Powell's model prediction, and the experimental data, respectively. The Mach number of the fully expanded jet flow is 1.49. }
		\label{1.49_1}
	\end{figure}

To examine the prediction at different operation conditions, we also compare the screech directivity patterns with the C mode observed in \citet{1983Norm}. In this case, the Mach number of the fully expanded jet flow is 1.49. The corresponding azimuthal mode of instability waves is also $n=1$.  Similar to Fig. \ref{1.19_1}, we show the experimental data, this model's prediction, and Powell's model prediction in Fig.~\ref{1.49_1}. Again, the sound radiation due to one-cell interaction predicted by Powell's model is uniform in every direction. Therefore, we do not repeat the description for brevity.

Regarding the fundamental tone, as can be seen in Fig.~\ref{1.49_1}(a), the present model predicts a strong upstream lobe and a weaker downstream lobe, which aligns well with experimental lobe positions and relative intensities, although this is only due to one-cell interaction. When three sources are combined, as shown in Fig. \ref{1.49_1}(b), the prediction from the monopole array theory shows two major lobes of nearly equal strength, peaking at $0^\circ$ and $180^\circ$, respectively. The relative intensity of the two lobes compares less favorably with the experimental data. In particular, the predicted radiation monotonously increases as $\psi$ approaches $180^\circ$, which clearly contradicts the experimental results. However, the present model shows much better agreement with the experimental data regarding the lobe position, width and relative intensity. As expected, the quick decay in the upstream and downstream directions is captured by the present model.

For the first harmonic, as shown in Fig. \ref{1.49_1}(c), the present model predicts four lobes peaking around $165^\circ$, $110^\circ$, $75^\circ$, and $30^\circ$, respectively. Experiments show three similar lobes, two major lobes pointing to $160^\circ$ and $90^\circ$, respectively, and a third lobe with a much smaller intensity peaking at $30^\circ$. The overall shape 
of the experimental data generally follows that of the prediction. When considering three sources, as shown in Fig. \ref{1.49_1}(d), the monopole array theory predicts three lobes of nearly the same intensity ($0^\circ$, $85^\circ$, and $180^\circ$). The intensity of the lobe in the downstream direction is overpredicted. In contrast, the present model predicts two major lobes and one minor lobe, whose width and peaking angle agree better with the experiment. Note that the experiment exhibits a rapid decay as $\psi$ approaches $180^\circ$. This is successfully captured by the present model. However, the amplitude of the downstream lobe is also overpredicted. The reason is not yet clear.

As for the second harmonic, four lobes appear in our prediction, three dominant lobes peaking around $130^\circ$, $95^\circ$, and $70^\circ$, respectively, while a much weaker lobe radiates to $20^\circ$. Experiments show one principal lobe radiating upstream,  two secondary lobes to the side of the jet ($90^\circ$), and a minor lobe around $30^\circ$. The lobe positions of the experimental data exhibit behavior similar to that predicted by the model. When the three sources are combined, the present and Powell's models yield similar agreement with the data, apart from the better agreement close to $180^\circ$ in this model.

\begin{figure}
	\centering
	\includegraphics[width = 0.75\textwidth]{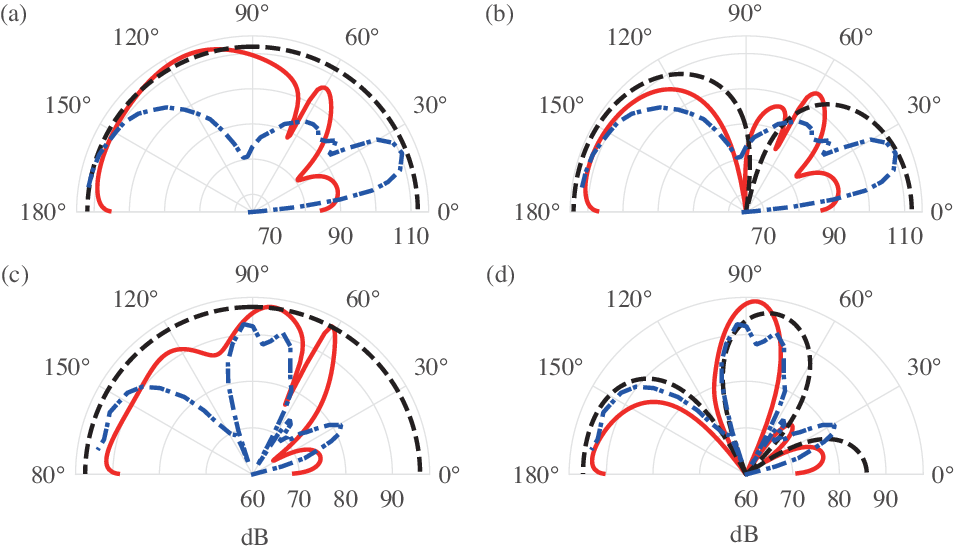}
	\caption{Comparisons of the far-field directivity patterns obtained by our model, by Powell's model, and those extracted from~\citet{direcity_numerical_tam}: Panels $(a, b)$ the fundamental frequency and $(c, d)$ the first harmonic. The red solid lines, the black dashed lines, and the dot-dashed lines represent this model's prediction, Powell's model prediction and the numerical data, respectively. The Mach number of the fully expanded jet flow is 1.18.}
		\label{tam_solo}
	\end{figure}

In addition to experimental results, we also include comparisons with numerical simulations \citep{direcity_numerical_tam}, where an axisymmetric screeching jet was simulated with $M_{-}=1.18$. The numerical results, this model's prediction, and Powell's model prediction are shown in Fig. \ref{tam_solo}. 
For the fundamental tone, this model predicts three lobes as shown in Fig.~\ref{tam_solo}(a). A dominant lobe radiates to the upstream direction and two secondary lobes peak
around $60^\circ$ and $20^\circ$, respectively. The numerical results exhibit two major lobes, one in the upstream direction and the other in the downstream direction. In addition, another lobe with a much smaller intensity peaks around $\psi=60^\circ$. The agreement is good between this model's prediction and the numerical simulations in terms of the lobe position, whereas the relative intensity and lobe width do not match well. When considering three sources, Powell's model predicts two lobes with nearly equal intensity, peaking at $180^\circ$ and $0^\circ$, respectively.  The secondary lobe peaking at $60^\circ$ shown in experiments appears not to be predicted. In contrast, the present model predicts four lobes, one major lobe to the upstream direction, while three minor lobes peaking at $15^\circ$, $60^\circ$, and $75^\circ$, respectively. The intermediate two lobes seem to match those in the numerical simulation. However, similar to Powell's model prediction, the relative intensity of these lobes does not match well with the numerical data. 

As for the first harmonic, the present model predicts three dominant lobes (peaking at $150^\circ$, $80^\circ$, and $60^\circ$) and a minor lobe in the downstream direction. In contrast, the numerical results have two major lobes (peaking at $150^\circ$ and $90^\circ$). The position of these two lobes is in good agreement with that in the prediction. In addition, a minor lobe appears in the downstream direction. The location of this lobe is in satisfactory agreement with that in the prediction. Interestingly, we see a significantly weak lobe in the numerical data at $60^\circ$. This agrees well with a major lobe in the prediction in terms of lobe position, albeit at a much weaker intensity. When three sources are considered, as shown in Fig.~\ref{tam_solo}(d), we see that the two models predict similar directivity patterns. However, the present model better agrees with the numerical data both in terms of lobe width and relative intensity.  It is particularly the case for the minor lobe appearing at around $30^\circ$, where the rapid decay when $\psi$ approaches $0^\circ$ is also captured.

In summary, it can be seen that the sound source due to one-cell interaction is far from of a monopole type. Instead, it exhibits similar characteristics to the experimental and numerical data in terms of the overall shape.  After incorporating multiple shock-cell interactions, the resulting sound directivity patterns show better agreement with the experimental data than those from the monopole array theory, in terms of both the lobe position and its relative intensity. In particular, unlike the monopole array theory, the present model can capture the rapid decay of the noise intensity as observer angles approach $0^\circ$ and $180^\circ$. We conclude that the distinct directivity patterns of screeching tones do not result from pure constructive interference. Instead, the sound source generated from the one-cell interaction can reveal many essential features shown in the experimental data. Nevertheless, constructive interference plays a dominant role in determining the final directivity pattern. After incorporating multiple shock-cell interactions, the far-field directivity patterns can be predicted more accurately.  

\subsubsection{Effects of the spatial growth rate of instability waves on
directivity patterns }
\label{subsub:effects}

\begin{figure}
    \centering
    \includegraphics[width = 0.92\textwidth]{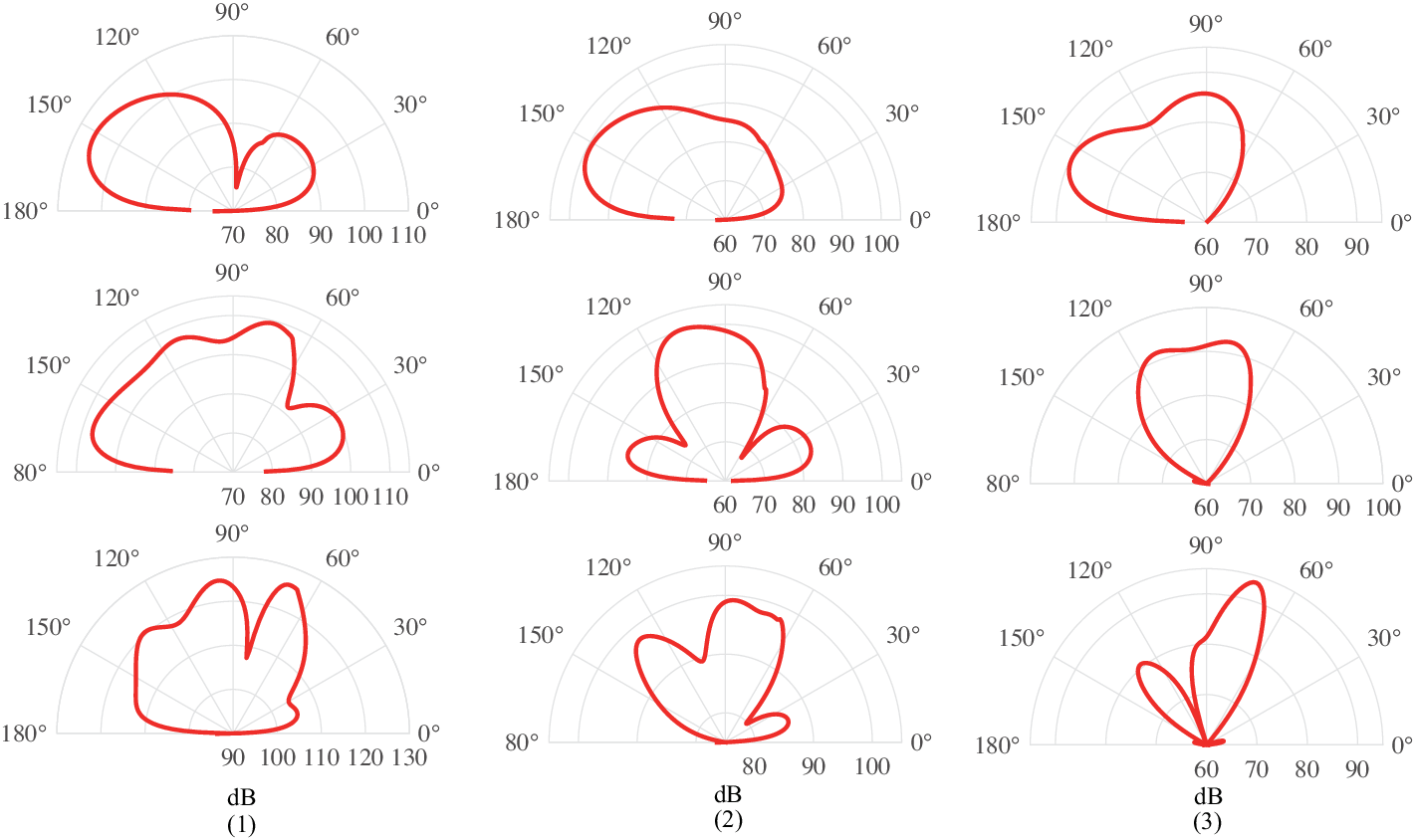}
    \caption{Effects of the spatial growth rate of instability waves on
	directivity patterns. The imaginary parts of the instability wave
	wavenumber are $\alpha_{i}, \alpha_{i}/2$, and 0 for (1), (2), and (3),
	respectively. Other operation conditions are kept the same as those used
    in Fig. \ref{1.49_1}. }
    \label{effects_on_directivity}
\end{figure}
The spatial wavenumber of the instability wave is calculated via the dispersion
relation for a vortex sheet in this model. Therefore, the instability wave grows
exponentially inside one shock structure. However, it is well-established that
nonlinear effects become significant as the instability waves grow to large
amplitudes. When screech occurs, the instability wave may grow marginally or
even saturate, rather than continuing to grow exponentially. Although this
linear model cannot capture such nonlinear behavior, we can still adjust the
spatial growth rate of the instability waves, i.e. the imaginary part of the
wavenumber, to assess its impact on the directivity patterns.

In our previous work \citep{li_and_lyu}, we found that reducing the spatial
growth rate of instability waves in two-dimensional jets reduces the sound
radiated in the downstream direction. In a similar manner, for this round jet,
we reduce $\alpha_i$ first to half its original value and then to zero in order to examine
its effect on the directivity patterns. The results are shown in
Fig.~\ref{effects_on_directivity}, where cases (1)-(3) correspond to imaginary
parts of the instability wave wavenumber equal to $\alpha_i$, $\alpha_i/2$, and
0, respectively. First, we see that the overall intensity of the generated sound
decreases as the imaginary part decreases, which is expected because the growth rate of the instability waves is
reduced. Second, similar to that in the two-dimensional case, the
downstream-directed sound generally diminishes from (1) to (3). Specifically,
when $\alpha_i = 0$ (case 3), virtually no sound is radiated in the direction
$\psi<60^\circ$.

 \add{Building on the two-dimensional study, which reported only a decrease in downstream radiation
when $\alpha_i$ was reduced, we observe here a uniform reduction across all the
side lobes. Consequently, the directivity patterns become spatially more
confined, collapsing into a primary lobe at $150^\circ$ for the fundamental
tone, $90^\circ$ for the first harmonic, and $75^\circ$ for the second
harmonic (a secondary lobe also can be seen at $130^\circ$ but with a much weaker intensity compared to the major one).} 
 Similar trends are
observed under other operation conditions (omitted here for brevity). It appears
that when $\alpha_i$ increases, acoustic emission occurs in an increasingly
large range of observer angles. This behavior is consistent with the Mach wave
radiation as described by \citet{1995_annu_tam}: for a single
wavenumber, only one phase speed exists, and Mach waves radiate in a single
direction. However, the growth and decay of the instability wave amplitude
introduce a broadband wavenumber spectrum, resulting in Mach wave radiation over
a wider range of angles. In our round‑jet model, increasing the imaginary
component of the instability wavenumber similarly broadens the spectrum, and
hence produces a correspondingly wider acoustic radiation pattern. 

\subsection{Sound generation efficiency spectra due to the SII} 

\label{subsec:spectra} 

\co{As noted in Sec.\ref{Sec:introduction}, in addition to screech,
Eq.(\ref{steepest_decent_way_P}) may also be used to model the BBSAN. However,
to predict its far-field spectra, a known spectral amplitude of the instability
wave is required. It is often difficult to obtain such an amplitude without
performing high-fidelity numerical simulations. Nevertheless, one can use
Eq.(\ref{steepest_decent_way_P}) to examine the efficiency of sound generation due
to the SII at various frequencies. Note that, if a uniform amplitude is assumed for
the instability waves at all frequencies, the sound generation efficiency also
represents the far-field spectra of the BBSAN. Even in the case where the spectrum
is not uniform, one expects that such a study may be able to capture local spectral
features for the BBSAN, particularly when the sound generation efficiency
exhibits marked variations when the frequency varies.}
\begin{figure}
    \centering
    \includegraphics[width = 0.7\textwidth]{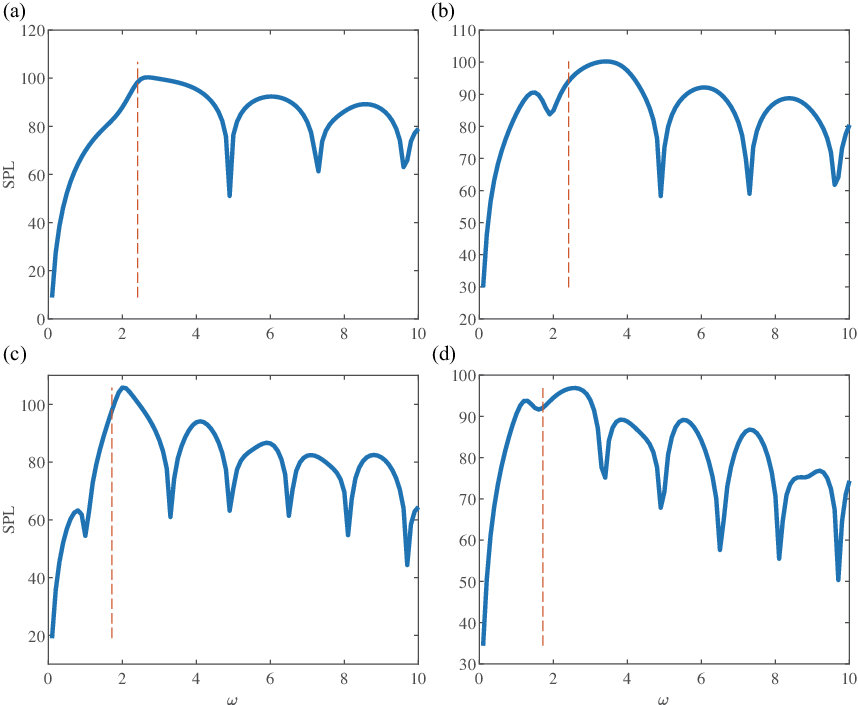}
    \caption{Sound generation efficiency spectra due to the SII at the observer
    angle $\psi=150^\circ$. The jet Mach numbers for panels $(a, b)$ and
$(c, d)$ are 1.3 and 1.5, respectively. The azimuthal modes of the acoustic wave
are $n=1$ and $n=0$ for panels $(a, b)$ and $(c, d)$, respectively.}
    \label{fig:spectra}
\end{figure}

\co{We have shown that the growth rate of the instability waves affects the
acoustic field generated by the SII considerably. Given that the BBSAN is
likely to be dominated by the interaction between the instability waves of the
largest amplitudes and shock-cell structures, it appears reasonable to assume a
saturated amplitude for the instability waves. Therefore,  in the following, we
choose to retain only the real part of $\alpha$ computed from
Eq.~(\ref{steepest_decent_way_P})}. \co{In Fig. \ref{fig:spectra}, we show the efficiency
spectrum of SII sound generation in terms of SPL defined by Eq. (\ref{equ:SPL}).}

\begin{figure}
    \centering
    \includegraphics[width = 0.7\textwidth]{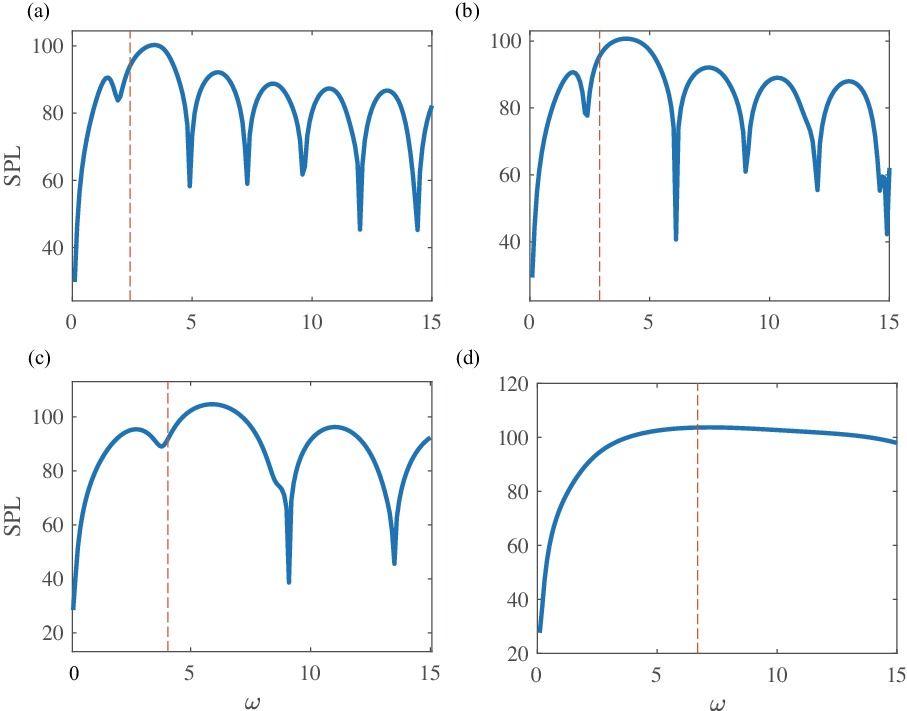}
    \caption{Sound generation efficiency spectra due to the SII when $M_-=1.3$ and $n=0$. The observer angle for panels (a)-(d) are $150^\circ$, $120^\circ$, $90^\circ$, and $60^\circ$, respectively.} 
    \label{fig:spectra_2}
\end{figure}

As shown in Fig. \ref{fig:spectra}, under each operation condition, the SPL
rapidly increases from $\omega=0$ to the peak frequency and then generally
decreases when $\omega$ further increases. \add{A pronounced bulge} \add{ is
evident in the spectra. Since the variation is rapid, one expects that such a bulge
may still be present in the BBSAN spectrum. In fact, this accords to the spectral
bulge observed in experimental BBSAN spectra. Additionally, quasi-periodic
structures are observed \co{in Fig. \ref{fig:spectra}}. } Both the bandwidth and the peak
frequency of the BBSAN bulge decrease when increasing the jet Mach number, as
seen by comparing panels (a, b) to (c, d). In contrast, varying the azimuthal
mode from $n = 1$ to $n = 0$ does not appear to introduce significant changes in the spectral
shape.

According to Tam \citep{tam_machwave}, 
the peak
frequency of the BBSAN's spectral bulge is given by
\begin{equation}
\omega_m=\frac{2\pi/S}{1/\kappa-M_+ \cos\psi}.
\label{equ:BBSANomega_m}
\end{equation}
The red dashed lines in Fig.\ref{fig:spectra}, computed from
Eq.(\ref{equ:BBSANomega_m}), show satisfactory agreement with the predicted peak
frequencies. In Fig.\ref{fig:spectra_2}, we further vary the observer angle while keeping the jet Mach number at 1.3 and the azimuthal mode at 0. The predictions by Eq.(\ref{equ:BBSANomega_m}) remain within the bulge corresponding to the BBSAN. Additionally, the bandwidth broadens when the observer angle shifts toward the downstream direction.
Such predicted BBSAN bandwidth quantitatively reproduces the experimental
measurements reported by \citet{1986Tam_Proposed}.  \add{We see that although
our model is based on the vortex-sheet assumption, it still captures several key
spectral features of the BBSAN observed in experiments. Therefore, one may
expect a better agreement if an accurate spectral
amplitude of instability waves obtained from experiments or numerical
simulations is used.}


\subsection{Near-field pressure and noise generation mechanism}
\label{sec:near field}

\begin{figure}
    \centering
    \includegraphics[width =0.95\textwidth]{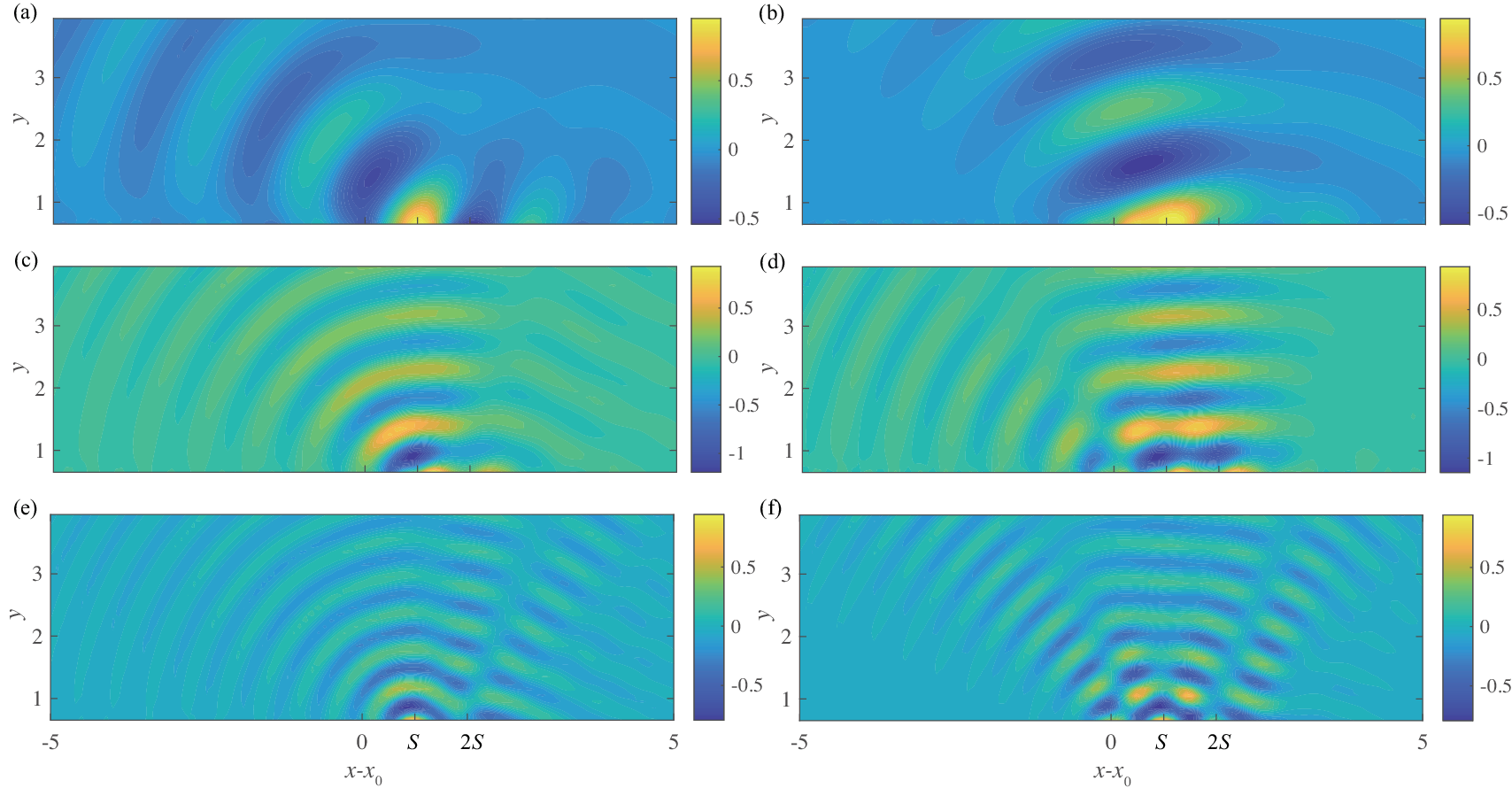}
    \caption{The normalized near-field pressure immediately outside the jet~($x_{0}$ is the starting position of the effective source for one-cell interaction):  $(a, b)$ the fundamental frequency; $(c, d)$ the first harmonic; $(e, f)$ the second harmonic. Panels $(a, c, e)$ consider one-cell interaction, while $(b, d, f)$ are for three cell interactions. The intensity of the three sources are kept as (0.5, 1, 0.5).}
    \label{fig:spl_1.5}
\end{figure}

In order to understand the physical mechanism of the noise generation, we further examine the near-field pressure distribution in the vicinity of the jet due to the SII. To obtain such a solution, Eq.(\ref{equ:numerical integration}) is integrated numerically. Fig.~\ref{fig:spl_1.5} shows the results when the Mach number of the fully expanded jet flow is 1.19. Considering one-cell interaction (in panels (a, c, e)), we can find that
 the near-field pressure mainly radiates to the upstream direction and has a weaker distribution downstream of the jet at the fundamental tone. In contrast, at its first harmonic, the near-field pressure shows a main lobe to the side of the jet ($90^\circ$), besides which a weaker lobe pointing to the downstream direction can be observed. The peaking angles are in good agreement with those of the far-field directivity pattern shown in Fig.~\ref{1.19_1}. Similar agreement between the near- and far-field is achieved for the second harmonic, and in the case of the three-cell interactions, we do not repeat the description for brevity.

\add{Fig. \ref{fig:spl_1.5} also shows the locations of the shock cells used to
calculate the near‐field perturbations, indicated by the short black ticks on
the horizontal axis. For the one–cell interaction, the effective source region
extends from $0$ to $S$, whereas for the three–cell interactions it spans from
$-S$ to $2S$. In the one–cell interaction, the pressure perturbations seem to be
concentrated near the end of the shock structure (the shock tip), as illustrated
in Fig.~\ref{fig:spl_1.5}(a), (c), and (e). For the three source configuration,
the harmonics exhibit a similar localization (see Fig. \ref{fig:spl_1.5}(d) and
(f)). In contrast, the fundamental tone appears to originate from the entire
shock structure over the interval $[0,S]$.  This behavior may be due to the
relatively large fundamental wavelength, which causes multiple sources to merge
into one effective region. }Note that the shock tip seems to serve as an
effective source location for screech. This can be attributed to the growth of
instability waves along the streamwise direction, which causes the source term
in the inhomogeneous wave equation (\ref{equ_sound_inhomogenous2}) to reach its
maximum at the shock tip. Such localised source regions have been reported in
several experimental studies and may be correlated with the spatial distribution
of instability‐wave amplitude as indicated by this model.

Based on the present model and the computed near-field pressure distributions, we are now in a position to investigate the mechanism of noise generation due to the SII. As indicated in Sec. \ref{subsec:spectra}, the peak frequency in the sound generation efficiency spectra agrees well with that predicted by Eq. (\ref{equ:BBSANomega_m}), which is derived from the Mach wave radiation mechanism. According to~\citet{tam_machwave},
in supersonic jets, the near-field pressure fluctuations can convect at speeds greater than the ambient speed of sound, thus enabling Mach wave radiation. The classical Mach angle satisfies $\psi^* = \arccos  ({c_0}/{u_x}),$
where $u_x = -\omega/\lambda$ denotes the phase velocity of the near-field pressure perturbations, and $c_0 = 1/M_+$ represents the ambient sound speed. Substituting these relations into the original relation yields
\begin{equation}
    \psi^* = \arccos \left( -\frac{\lambda}{\omega M_+} \right).
    \label{equ:Machradiation}
\end{equation}

In our model, the acoustic velocity potential can be expressed as an integral over the streamwise wavenumber $\lambda$ (see (Eq.\ref{equ:far-field estimation})).
Here, $D_{+}(\lambda)$ denotes the spectral term arising from the SII (see the source term of Eq.~(\ref{equ_sound_inhomogenous2})), which is directly related to the Fourier transform of the near-field pressure along a fixed radial position. In the far field,
 Eq. (\ref{equ:far-field estimation}) can be evaluated using the steepest descent method. The saddle point $\lambda_0$ associated with a given observer angle $\psi$ satisfies
\begin{equation}
    \lambda_0 = -\omega M_+ \cos \psi.
    \label{equ:saddle_point}
\end{equation}
Physically, this means that $\lambda_0$ corresponds to the streamwise wavenumber component of the sound wave propagating along the direction $\psi$. 

\begin{figure}
    \centering
   \includegraphics[width = 0.5\textwidth]{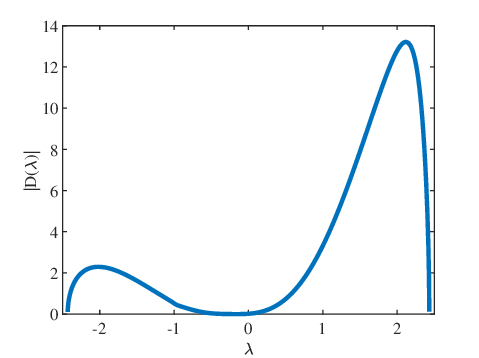}
   \caption{$|D(\lambda)|$ at the frequency of the fundamental tone. The Mach number of the fully expanded jet is 1.49.}
    \label{fig:mechanism_3}
\end{figure}

Form Eqs. (\ref{equ:Machradiation}) and (\ref{equ:saddle_point}), it is straightforward to see that the saddle point $\lambda_0$ corresponds to the streamwise wavenumber component of the Mach wave radiated at $\psi^*$. As the observer angle $\psi^*$ varies from $0$ to $\pi$, the saddle point $\lambda_0$ ranges accordingly from $-\omega M_+$ to $\omega M_+$. 
Thus, the far-field radiation pattern can be interpreted by examining the distribution of the spectral coefficient $|D_+(\lambda)|$ over this interval.
Since each observer angle corresponds to a specific $\lambda_0$, the magnitude $|D_+(\lambda_0)|$ effectively determines how much acoustic energy is radiated in that direction.
As shown in Fig.~\ref{fig:mechanism_3}, the quantity $|D(\lambda)|$ exhibits two prominent peaks at $\lambda < 0$ and $\lambda > 0$, corresponding to radiation angles $\psi \approx 148^\circ$ and $47^\circ$, respectively, for a fully expanded Mach number $M_- = 1.49$. These predictions are in excellent agreement with the directivity patterns shown in Fig.~\ref{1.49_1}(a), confirming that the dominant acoustic radiation is governed by the Mach wave radiation mechanism.

In summary, these results support the following physical picture: near-field pressure fluctuations caused by the SII travel with supersonic phase speeds. When the streamwise wavenumber of these fluctuations matches that of a freely propagating acoustic wave in the ambient field, energy is radiated in the corresponding direction. The spectral shape of $|D_+(\lambda_0)|$, governed by the SII, determines the primary lobes of the acoustic directivity patterns.

\section{Discussion}
\label{sec:discussion}
\subsection{Comparison between round and two-dimensional jets}
According to Sec. \ref{section:results and discussion} and \citet{li_and_lyu},
we see that acoustic emission due to the interaction between shock and
instability waves shares some essential similarities between round and
two-dimensional jets. First, the sound source resulting from one-cell
interaction in both round and two-dimensional jets is not of the monopole type
assumed by Powell. Instead, it captures many essential features of the screech's
directivity patterns. Second, the predicted directivity patterns from both
models show good agreement with the experimental data after incorporating
multiple shock-cell interactions, capturing the rapid decay as the observable
angle approaches $180^\circ$. Third, the noise generation mechanism in both
cases can be explained by the Mach wave mechanism, \add{ and the directivity
patterns become more concentrated into primary lobes as the imaginary part of
$\alpha$ is reduced.}
\add{Another interesting observation is that, in both round and two-dimensional
jets, the directivity of the first harmonic predicted by the multiple shock cell
model shows good agreement with the experimental data. }


However, some important differences also exist. For example, the complementary solutions of the convective-wave equation that governs the generated acoustic waves take different forms. Specifically, the acoustic waves outside the jet can be respectively written as
\begin{equation}
 		\frac{1}{2\pi}\int_{-\infty}^{+\infty}D^{d}_{+}(\lambda)\mathrm{e}^{-\mathrm{i}(\lambda x-\gamma_{+}y)}\mathrm{d}\lambda,\quad \frac{1}{2\pi}\int_{-\infty}^{+\infty}D_{+}(\lambda)H_{n}(\gamma_{+}r)\mathrm{e}^{-\mathrm{i}\lambda x}\mathrm{d}\lambda.
		\label{equ:two_near}
\end{equation}
Here $D^{d}_{+}(\lambda)$ is a coefficient similar to $D_{+}(\lambda)$ for two-dimensional jets and $y$ represents the cross-stream coordinate. The far-field sound can be estimated using the steepest descent method to evaluate Eq.(\ref{equ:two_near}), leading to results that are proportional to
\begin{equation} 
\frac{D^{d}_{+}(\lambda_0)}{\sqrt{R}}\sin \psi \quad \mathrm{and} \quad \frac{D_+(\lambda_0)}{R},
\label{equ:far_appro}
\end{equation}
respectively.
Eq.(\ref{equ:far_appro}) shows that the acoustic amplitude decays as $R^{-1/2}$ for two-dimensional jets and as $R^{-1}$ for round jets, which is a well-established fact. Additionally, the directivity patterns are solely determined by $|D_+(\lambda_0)|$ for round jets, while they are determined by  $|D^{d}_+(\lambda_0)\sin\psi|$ for two-dimensional jets. In two-dimensional jets, $|D^{d}_+(\lambda_0)|$ can capture the overall shape of the directivity pattern but reaches its maximum when $\psi$ approaches $180^\circ$. It is the factor $\sin\psi$ that ensure the rapid decay of the far-field sound when $\psi\rightarrow 180^\circ$.

\subsection{The role of the guided jet mode in screech feedback loop }
It is widely acknowledged that the guided jet mode plays a crucial role in
completing the screech feedback loop. Note, however, this model does not involve
guided-jet modes. Nevertheless, it captures many essential features of the
directivity patterns of screech. This suggests that, although guided jet modes
play an important role in closing the feedback loop and determining the screech frequency, the directivity of
the sound field is primarily governed by the interaction between the instability
waves and the shock-cell structures.

\section{Conclusion} \label{sec:conclusion}
In this paper, an analytical model is developed to predict the sound generated
by the interactions between shock and instability waves in supersonic round
jets. The jet is assumed to be of the vortex-sheet type. Both shock and
instability waves are assumed to be of small amplitudes so Euler equations are
linearized to determine the governing equations for shock, instability, and
their interaction, respectively. The resulting sound is obtained by solving an
inhomogeneous wave equation while simultaneously matching kinematic and dynamic
conditions on the vortex sheet. The directivity patterns of screeching tones are
obtained by the saddle point method in the far field.

We first concentrate on one-cell interaction and subsequently incorporate
multiple shock-cell interactions. Results show good agreement with the
experimental and numerical data in terms of both the lobe's position and its
relative intensity, even when only one-cell interaction is considered. We may,
therefore, conclude that the distinct directivity patterns of screeching tones
do not result from pure constructive interference. Instead, the sound source
generated from the one-cell interaction can reveal many essential features shown
in the experimental data. Additionally, the present model correctly captures the
rapid decay of the acoustic radiation when $\psi$ approaches $180^\circ$ and
$0^\circ$. 
In particular, although the noise radiation primarily
occurs in the upstream direction at the fundamental frequency, it becomes weaker
as the observer angle approaches $180^\circ$, which better agrees with
experimental results compared with earlier models. \add{We also investigate the
influence of the instability wave growth rate on the directivity patterns. We
find that the amplification and decay of the instability wave amplitude produce
a broadband wavenumber spectrum, which in turn generates acoustic radiation over
a wider range of observer angles.} 
 
\add{Subsequently, we examine the sound generation efficiency spectra 
due to
the SII. The peak frequency matches well with that predicted by the Mach wave
radiation mechanism. Moreover, the predicted broadband noise from the SII
qualitatively agrees with some experimental observations.}

Finally, we examine the near-field pressure distribution in order to identify the sound generation mechanism for screech. We find that the near-field pressure distributions are consistent with those in the far-field directivity patterns.
By examining the wavenumber matching of the near-field pressure, we find that screech is generated primarily through the Mach wave mechanism. An ``effective source location" may be identified at the tip of the shock cell, which can be explained by the growth of instability waves.

While this study focuses on shock-instability interaction mechanisms in isolated jets, future work can extend to installed configurations. Recent models suggest that jet-surface interactions (e.g., trailing-edge scattering of near-field waves and surface reflection) dominate installed jet noise at low frequencies and amplify high-frequency noise on the reflected side \citep{lyu20177, Lyu_Dowling_2019_2, Lyu_Dowling_Naqavi_2017}. Such extensions can bridge intrinsic SII physics with installation effects relevant to aircraft design.

\section*{Declaration of competing interest}
The authors declare that they have no known competing financial interests or personal relationships that could have appeared
to influence the work reported in this paper.
\section*{Acknowledgments}
The authors wish to gratefully acknowledge the National Natural Science Foundation of China (NSFC) under the grant number 12472263. The second author (BL) wishes to acknowledge the funding from the Beijing Natural Science Foundation (L253027) and from Laoshan Laboratory (LSKJ202202000). 
\appendix
\section{}
\label{appendix1}

 Considering the boundary conditions at $r=0$ and $r\rightarrow\infty$, the velocity potential of the instability-induced perturbations can be written as
\begin{equation}
    \phi_{v}=
    \begin{cases}
       H_{n}(\mu_{+}r)\mathrm{e}^{{\mathrm i}(n\theta+\alpha x-\omega t)},& r> \frac{1}{2}, \\[2pt]
       \mathcal{C} J_{n}(\mu_{-}r)\mathrm{e}^{{\mathrm i}(n\theta+\alpha x-\omega t)},& r\leq \frac{1}{2},
    \end{cases} 
    \label{equ:instability_potential}
\end{equation}
where coefficient $\mathcal{C}$ can be obtained as
\begin{equation}
    \mathcal{C}=\frac{\omega-\alpha}{\omega}\frac{\mu_{+}}{\mu_{-}}\frac{H^\prime _{n}(\mu_{+}/2)}{J^\prime _{n}(\mu_{-}/2)}.
\end{equation}

Inside the jet, the instability-induced pressure and velocity perturbations can be written as
\begin{subequations}
\begin{align}
	p_{v-}=\mathrm{i}\frac{M_-^2}{M_+^2}\mathcal{C}(\omega-\alpha)J_n(\mu_- r) &, \\
    u_{v-}=\mathrm{i}\alpha\mathcal{C}(\omega-\alpha)J_n(\mu_- r)  &,\\
   v_{v-}=\mathcal{C}\mu_-   J_n^\prime(\mu_- r) &,\\
   w_{v-}=\mathrm{i}n\mathcal{C}J_n(\mu_- r)&.
 \end{align}
\end{subequations}
Outside the jet, the corresponding pressure and velocity perturbations read
\begin{subequations}
\begin{align}
	p_{v+}=\mathrm{i}\omega H_n(\mu_+ r) &, \\
    u_{v+}=\mathrm{i}\alpha H_n(\mu_+ r)  &,\\
   v_{v+}=\mu_+   H_n^\prime(\mu_+ r) &,\\
   w_{v+}=\mathrm{i}nH_n(\mu_+ r)&.
 \end{align}
\end{subequations}
The vortex sheet's displacement perturbation due to instability waves can be obtained as
\begin{equation}
h_v=\mathrm{i}\frac{\mu_+}{\omega}H_n^\prime (\mu_+ /2).
\end{equation}

The corresponding pressure, velocity, and vortex sheet's displacement perturbations due to shock structures take the following form,
\begin{subequations}
\begin{align}
	p_m=-\frac{M_-^2}{M_+^2} \frac{2\zeta \mathcal{A}}{\beta }J_{0}(2\zeta r)\cos(2\zeta x/\beta) &, \\
    u_m=\frac{2\zeta \mathcal{A}}{\beta }J_{0}(2\zeta r)\cos(2\zeta x/\beta) &,\\
   v_m=2\zeta\mathcal{A}J_{0}^\prime(2\zeta r)\sin(2\zeta x/\beta)  &,\\
   h_m=-\beta\mathcal{A}J_{0}^\prime(\zeta)\cos(2\zeta x/\beta)&.
 \end{align}
\end{subequations}
\section{}
\label{appendix2}
        According to \citet{li_and_lyu} and \citet{my_circular}, on the boundary
  $r =1/2$, the dynamic boundary
  condition for the total pressure reads
   \begin{equation}
      p_{i+}+h_{m} \frac{\partial p_{v+}}{\partial r }= p_{i-}+h_{m} \frac{\partial p_{v-}}{\partial r }+h_{v} \frac{\partial p_{m-}}{\partial r },
  \end{equation}
  while the kinematic boundary condition of the total displacement requires
  \begin{subequations}
    \begin{align}
         v_{i+}+\frac{\partial v_{v+}}{\partial r }h _{m}&=\frac{\partial h_{i} }{\partial t }+u_{v+}\frac{\partial h_{m} }{\partial x },\\
           v_{i-}+\frac{\partial v_{v-}}{\partial r }h _{m}+\frac{\partial v_{m-}}{\partial r }h _{v}&=\frac{\partial h_{i} }{\partial t }+U\frac{\partial h _{i}}{\partial x }+u_{v-}\frac{\partial h_{m} }{\partial x }+u_{m-}\frac{\partial h_{v} }{\partial x }.
    \end{align}     
  \end{subequations}
Here $h_m$ and $h_v$ represent the vortex sheet displacement induced by the shock and instability waves, respectively, which can be readily calculated from 
\begin{equation*}
    \frac{\mathrm{D}h_m}{\mathrm{D}t}=v_m\quad\mathrm{and}\quad \frac{\mathrm{D}h_v}{\mathrm{D}t}=v_v.
\end{equation*}

 Eq.(\ref{equ_sound_inhomogenous})  can be solved by assuming harmonic temporal and azimuthal variations $\mathrm{e}^{\mathrm{i}n\theta-\mathrm{i}\omega t}$.
 For brevity, we omit this term in what follows. Outside the jet, it is straightforward to see that the source term in Eq.(\ref{equ_sound_inhomogenous}) vanishes because there is no perturbation due to shock structures. In this case, Eq.(\ref{equ_sound_inhomogenous}) degenerates into a homogeneous
  wave equation, i.e.
    \begin{equation}
      \left[\frac{\partial^{2}}{\partial x^{2}}+\frac{\partial^{2}}{\partial r^{2}}+\frac{1}{r}\frac{\partial}{\partial r}-\left(\frac{n^{2}}{r^{2}}-\omega^{2}M_{+}^{2}\right)\right]\phi_{i+}=0.
      \label{equ_homogenous}
  \end{equation}
Inside the jet, $\phi_{i-}$ satisfies
\begin{equation}
    \begin{aligned}
        \left[\mathbf{\nabla}^{2}-M_{-}^{2}\left(-\mathrm{i}\omega +\frac{\partial}{\partial x}\right)^{2}\right]\phi_{i-}&=\left(\frac{2\zeta}{\beta}\frac{J_0}{J_0^\prime}\Xi_1 v_m+\Xi_2 u_m \right)u_v,\\
    \end{aligned}
    \label{equ_sound_inhomogenous2}
\end{equation}
where
\begin{subequations}
    \begin{equation}
    \begin{aligned}
     \Xi_1=-\frac{4\zeta}{\beta}-\frac{2\zeta}{\beta}\frac{\alpha-\omega}{\alpha}(\gamma-1)+2\beta\frac{\alpha-\omega}{\alpha}\frac{J_0^\prime}{J_0}\bigg[&\mu_{-}\frac{J_n^\prime}{J_n}+\\
    &(\gamma-1)\left(\frac{1}{2r}+\zeta\frac{J_0^{\prime\prime}}{J_0^\prime}\right)\bigg],
  \end{aligned}
  \end{equation}
  \begin{equation}    
  \begin{aligned}
    \Xi_2=2\mathrm{i}(\alpha-\omega)+\mathrm{i}\left(\alpha+\frac{n^2}{\alpha r^2}\right)(\gamma-1)+2\frac{\mu_{-}}{\mathrm{i}\alpha}&\frac{J_n^\prime}{J_n}\bigg[2\zeta\frac{J_0^\prime}{J_0}+\\
   & (\gamma-1)\left(\frac{1}{2r}+\frac{\mu_{-}}{2}\frac{J_n^{\prime\prime}}{J_n^\prime}\right)\bigg].
    \end{aligned}
     \end{equation}
\end{subequations}
Note that $J_{n}$ and $J_{0}$ represent $J_{n}(\mu_{-}r)$ and $J_{0}(2\zeta r)$, respectively.

With the harmonic temporal and azimuthal assumption, two boundary conditions can be rewritten as
\begin{subequations}
\begin{align}
    \omega \phi_{i+}-\frac{M_{-}^{2}}{M_{+}^{2}}\left(\omega+\mathrm{i}\frac{\partial}{\partial x}\right)\phi_{i-}=\mathcal{X}_{0}\cos\frac{2\zeta x}{\beta}\mathrm{e}^{\mathrm{i}\alpha x},&   \label{equ:boundary condition 1}\\
  \label{equ:boundary condition 2}
    \left(1+\frac{\mathrm{i}}{\omega}\frac{\partial}{\partial x}\right)\frac{\partial }{\partial r}\phi_{i+}-\frac{\partial }{\partial r}\phi_{i-}=\left(\mathcal{S}_{1} \sin{\frac{2\zeta x}{\beta}}
    +\mathcal{S}_{2}\cos\frac{2\zeta x}{\beta}\right)\mathrm{e}^{\mathrm{i}\alpha x},&
\end{align}
\end{subequations}
where
\begin{subequations}
\begin{align}
    \mathcal{X}_{0}=\mathcal{A}\beta J^\prime _{0}\left[\frac{M_{-}^{2}}{M_{+}^{2}}\mathcal{C}(\alpha-\omega)\mu_{-}J^\prime _{n}-\left(\frac{M_{-}^{2}}{M_{+}^{2}}\frac{4\zeta^{2}}{\beta^{2}}\frac{1}{\omega}-\omega\right)\mu_{+}H^\prime _{n}\right],&   \label{c_{1}}\\
    \mathcal{S}_{1}=2\mathrm{i}\mathcal{A}\zeta\left[\left(\alpha\left(1-\frac{\alpha}{\omega}\right) H_{n}-\frac{1}{\omega}\mu_{+}^{2}H^{\prime\prime}_{n}-\mathcal{C}\alpha  J_{n}\right)J^\prime _{0}+2\zeta\frac{1}{\omega}\mu_{+}H^{\prime}_{n}J^{\prime\prime}_{0}\right],&
    \label{c_{2}}\\
    \mathcal{S}_{2}=\mathcal{A}\left[\left(\beta\left(1-\frac{\alpha}{\omega}\right)\mu_{+}^{2}H^{\prime\prime}_{n}-\frac{4\zeta^{2}}{\beta}\frac{\alpha}{\omega}H_{n}-\mathcal{C}\beta\mu_{-}^{2}J^{\prime\prime}_{n}\right)J^\prime _{0}+\frac{2\zeta}{\beta}\frac{\alpha}{\omega}\mu_{+}H ^\prime_{n}J_{0}\right].&
 \label{c_{3}}
\end{align}
\end{subequations}
Note that $H_{n}$ represents $H_{n}(\mu_{+}/2)$ and $r=1/2$ in~Eq.(\ref{c_{1}})-(\ref{c_{3}}). 

\section{}
\label{appendixb}
For the following inhomogeneous equation, i.e.
\begin{equation}
    \left[\frac{\partial^{2}}{\partial r^{2}}+\frac{1}{r}\frac{\partial }{\partial r}-\left(\frac{n^{2}}{r^{2}}-k^{2}\right)\right]G=f(r,x),
\end{equation}
its particular solution takes the form as follows
\begin{equation}
    G^{p}(r,x)=\int\frac{J_{n}(ks)Y_{n}(kr)-J_{n}(kr)Y_{n}(ks)}{J_{n}(ks)Y^{\prime}_{n}(ks)-J_{n}^{\prime}(ks)Y_{n}(ks)}f(s,x) \mathrm{d}s.
    \label{equ:spetial}
\end{equation}
Note that
\begin{equation}
   J_{n}(ks)Y^{\prime}_{n}(ks)-J_{n}^{\prime}(ks)Y_{n}(ks)=\frac{2}{\pi ks}.
\end{equation}
Eq.(\ref{equ:spetial}) can be thus reorganized to
\begin{equation}
    G^{p}(r,x)=\frac{\pi k}{2}\int\big(J_{n}(ks)Y_{n}(kr)-J_{n}(kr)Y_{n}(ks)\big)s f(s,x)\mathrm{d}s.
\end{equation}
Therefore, after the Fourier transformation, the particular solution of Eq.(\ref{equ_sound_inhomogenous2}) can be written as
\begin{equation}
    \begin{aligned}
         G^{p}(r,\lambda)=\mathrm{i}\mathcal{C}\mathcal{A}\frac{\pi\zeta\alpha\gamma_{-}}{\beta}\int_{0}^{r}(J_{n}(\gamma_{-}s)Y_{n}(\gamma_{-}r)&-J_{n}(\gamma_{-}r)Y_{n}(\gamma_{-}s))s\\
     &J_0(2\zeta s)J_n(\mu_{-}s)\left[\Xi_{s1}\mathcal{I}_s+\Xi_{s2}\mathcal{I}_c\right]\mathrm{d}s,
     \label{equ:G_p}
    \end{aligned}
\end{equation}
where
\begin{equation*}
\begin{aligned}
    \Xi_{s1}=-\frac{4\zeta}{\beta}-\frac{2\zeta}{\beta}\frac{\alpha-\omega}{\alpha}(\gamma-1)+2\beta\frac{\alpha-\omega}{\alpha}&\frac{J_0^\prime(2\zeta s)}{J_0(2\zeta s)}\bigg[\mu_{-}\frac{J_n^\prime(\mu_{-}s)}{J_n(\mu_{-}s)}+\\
    &(\gamma-1)\left(\frac{1}{2s}+\zeta\frac{J_0^{\prime\prime}(2\zeta s)}{J_0^\prime(2\zeta s)}\right)\bigg],
\end{aligned}
\end{equation*} 
\begin{equation*}
    \begin{aligned}
          \Xi_{s2}=2\mathrm{i}(\alpha-\omega)+\mathrm{i}\left(\alpha+\frac{n^2}{\alpha s^2}\right)(\gamma-1)+2\frac{\mu_{-}}{\mathrm{i}\alpha}&\frac{J_n^\prime(\mu_{-}s)}{J_n(\mu_{-}s)}\bigg[2\zeta\frac{J_0^\prime(2\zeta s)}{J_0(2\zeta s)}+\\
          &(\gamma-1)\left(\frac{1}{2s}+\frac{\mu_{-}}{2}\frac{J_n^{\prime\prime}(\mu_{-}s)}{J_n^\prime(\mu_{-}s)}\right)\bigg].
    \end{aligned}   
\end{equation*}

\bibliographystyle{elsarticle-harv}

\bibliography{cas-refs}
\end{document}